\documentclass[journal]{vgtc}                     

\onlineid{1274}

\vgtccategory{IEEE VIS Full Paper}

\vgtcpapertype{Systems \& Rendering}

\title{TexGS-VolVis: Expressive Scene Editing for Volume Visualization via Textured Gaussian Splatting}

\author{%
  \authororcid{Kaiyuan Tang}{0009-0001-3512-0112},
  \authororcid{Kuangshi Ai}{0009-0005-7171-6529},
  \authororcid{Jun Han}{0000-0002-7286-062X}, and
  \authororcid{Chaoli Wang}{0000-0002-0859-3619}
}

\authorfooter{
  \item K. Tang, K. Ai, and C. Wang are with the Department of Computer Science and Engineering, University of Notre Dame, Notre Dame, IN 46556, USA.\\
E-mail: \{ktang2, kai, chaoli.wang\}@nd.edu.\\
  \item J.\ Han is with the Division of Emerging Interdisciplinary Areas and the Center for Ocean Research in Hong Kong and Macau (CORE), The Hong Kong University of Science and Technology, Hong Kong, China.\\ 
E-mail: hanjun@ust.hk.
}

\abstract{Advancements in volume visualization (VolVis) focus on extracting insights from 3D volumetric data by generating visually compelling renderings that reveal complex internal structures. Existing VolVis approaches have explored non-photorealistic rendering techniques to enhance the clarity, expressiveness, and informativeness of visual communication. While effective, these methods often rely on complex predefined rules and are limited to transferring a single style, restricting their flexibility. To overcome these limitations, we advocate the representation of VolVis scenes using differentiable Gaussian primitives combined with pretrained large models to enable arbitrary style transfer and real-time rendering. However, conventional 3D Gaussian primitives tightly couple geometry and appearance, leading to suboptimal stylization results. To address this, we introduce TexGS-VolVis, a textured Gaussian splatting framework for VolVis. TexGS-VolVis employs 2D Gaussian primitives, extending each Gaussian with additional texture and shading attributes, resulting in higher-quality, geometry-consistent stylization and enhanced lighting control during inference. Despite these improvements, achieving flexible and controllable scene editing remains challenging. To further enhance stylization, we develop image- and text-driven non-photorealistic scene editing tailored for TexGS-VolVis and 2D-lift-3D segmentation to enable partial editing with fine-grained control. We evaluate TexGS-VolVis both qualitatively and quantitatively across various volume rendering scenes, demonstrating its superiority over existing methods in terms of efficiency, visual quality, and editing flexibility.}


\graphicspath{{figs/}{figures/}{pictures/}{images/}{./}} 

\usepackage{tabu}                      
\usepackage{booktabs}                  
\usepackage{lipsum}                    
\usepackage{mwe}                       

\usepackage{mathptmx}                  
\usepackage{multirow}
\usepackage{graphicx}
\usepackage{times}
\usepackage{algorithm}
\usepackage{algorithmic}
\usepackage{amsfonts}
\usepackage{booktabs}         
\usepackage{gensymb}
\usepackage{amsmath}
\usepackage{comment}
\usepackage{times} 
\usepackage{caption}
\usepackage{bm}
\usepackage{multirow}
\usepackage{color} 
\usepackage{anyfontsize}
\usepackage{soul}

\usepackage{arydshln}

\DeclareMathOperator*{\nnfm}{NNFM}
\DeclareMathOperator*{\prompt}{prmt}
\DeclareMathOperator*{\img}{img}
\DeclareMathOperator*{\dep}{dep}
\DeclareMathOperator*{\ind}{ind}

\DeclareMathOperator*{\size}{size}
\DeclareMathOperator*{\total}{total}

\DeclareMathOperator*{\palette}{pal}
\DeclareMathOperator*{\tex}{tex}

\DeclareMathAlphabet{\altmathcal}{OMS}{cmsy}{m}{n}

\DeclareMathOperator*{\gt}{GT}

\newcommand{\hot}[1]{{\color{black} #1}}

\usepackage{caption}
\newenvironment{myitemize}{
\begin{itemize}
 \setlength{\itemsep}{1pt}
 \setlength{\parskip}{0pt}
 \setlength{\parsep}{0pt}}{\end{itemize}
 
}

\begin{document}

\firstsection{Introduction}
\maketitle
Throughout history, researchers—including biologists, surgeons, and physicochemical engineers—have sought to present the volumetric data they collect in a way that is both clear and accessible to others. 
{\em Volume visualization} (VolVis) plays a crucial role in scientific visualization, enabling researchers across various disciplines to effectively convey complex data and facilitate insightful analysis. 
One of the core techniques in VolVis is {\em direct volume rendering} (DVR).
DVR utilizes a {\em transfer function} (TF) to map each voxel to a specific color and opacity based on its value. 
It generates visualization images that reveal internal structures, highlight essential features, and enable in-depth exploration of complex volumetric datasets.

\hot{However, visualization using TFs is inherently limited in expressiveness, offering minimal control over visual style, texture semantics, or perceptual emphasis beyond basic color-opacity mappings. 
As a result, traditional DVR often struggles to convey complex structures, subtle variations, or high-level semantics in a visually compelling way. 
To enhance the expressiveness of VolVis for both artistic and analytical purposes, prior studies within the visualization community~\cite{Laidlaw-VIS98, Lu-VIS02, Tang-VIS24} have explored the integration of {\em non-photorealistic rendering} (NPR) techniques to emulate artistic styles.} 
For instance, Laidlaw~\cite{Laidlaw-VIS98} used expressive and discrete strokes inspired by van Gogh's paintings to visualize multidimensional data.
Volume stippling~\cite{Lu-VIS02} achieves illustrative volume rendering by mimicking the style of stipple drawings. 
Traditional NPR methods, however, rely on predefined, complex generation rules to simulate specific artistic styles, limiting each method to a single style. 
A more recent approach, StyleRF-VolVis~\cite{Tang-VIS24}, tackled this issue by employing a {\em neural radiance field} (NeRF) to reconstruct the VolVis scene, followed by the use of a pre-trained VGG network to transfer style from a reference image to the VolVis scene's appearance. 
Despite its ability to transfer arbitrary styles, this framework has several practical limitations. 
First, as a {\em novel view synthesis} (NVS) method, StyleRF-VolVis reconstructs the VolVis scene from multi-view images rendered with a specific TF. 
This approach limits its ability to represent or stylize the invisible parts of the scene (i.e., regions with zero opacity in the TF). 
Second, the neural network architecture in StyleRF-VolVis is computationally expensive to train and render, leading to inefficient style transfer and low rendering frame rates. 
Finally, StyleRF-VolVis requires a reference style image containing the desired artistic style. 
However, finding such an image can be challenging; in some cases, a suitable reference may not exist.

To address these limitations, we introduce TexGS-VolVis, a {\em Gaussian splatting} (GS) method designed for expressive VolVis scene representation and editing. 
Unlike recent 3DGS methods~\cite{Niedermayr-arxiv24, Tang-PVIS25} for VolVis scene representation, TexGS-VolVis uses 2DGS~\cite{Huang-SIGGRAPH24} as its backbone and extends each primitive with an additional texture attribute. 
By combining the geometric reconstruction strengths of 2DGS with the flexibility of the texture attribute, TexGS-VolVis enables a more accurate reconstruction of the underlying VolVis scene geometry. 
This decouples the appearance representation of each Gaussian primitive from its geometric expression, facilitating more flexible and high-quality {\em non-photorealistic scene editing} (NPSE). 
\hot{When optimizing NVS models using a preset TF, regions mapped to zero opacity are invisible during training and therefore cannot be reconstructed.
To address this limitation, we leverage the {\em composability} of the Gaussian representation.
We optimize multiple {\em basic models}, each trained on a {\em basic scene} defined by a {\em basic TF} that covers a distinct, non-overlapping region of the volume.
By composing the parameters from all basic models, we construct a single composed model that jointly represents all regions of interest across the individual scenes.
}
Furthermore, TexGS-VolVis enables significantly faster rendering compared to StyleRF-VolVis by rasterizing differentiable Gaussian primitives directly onto the image plane, bypassing the need for a neural network feed-forward process.

For scene editing, we augment the Gaussian primitives with Blinn-Phong shading attributes in TexGS-VolVis, resulting in a VolVis scene representation model that supports {\em photorealistic scene editing} (PSE). 
This includes adjustments to color, opacity, and lighting, enhancing user exploration. 
Unlike StyleRF-VolVis, which is limited to image-driven NPSE, TexGS-VolVis offers more flexible scene editing methods. 
By leveraging the capabilities of large pretrained models~\cite{Kirillov-ICCV23, Radford-CLIP, Rombach-SD, brooks-InstructPix2Pix} developed in the computer vision field, TexGS-VolVis supports both image-driven and text-driven NPSE.
Furthermore, taking advantage of TexGS-VolVis's explicit representation, we employ a {\em 2D-lift-3D segmentation} approach. 
This allows users to segment different parts of a basic scene using a 2D segmentation model~\cite{Kirillov-ICCV23} and apply distinct editing effects to the corresponding Gaussian primitives. 
This enables partial stylization within a single basic scene, offering more granular control over the editing process. 
To summarize, we present the following contributions:
\begin{myitemize} 
\vspace{-0.05in}
\item We develop TexGS-VolVis, the first 2DGS-based model designed for expressive VolVis scene representation and editing. 
\item \hot{TexGS-VolVis integrates Blinn-Phong shading and texture attributes, supporting real-time PSE and high-quality NPSE.}
\item We leverage the power of existing large pre-trained models to enable flexible editing, including image-driven and text-driven NPSE and 2D-lift-3D scene segmentation. 
\item Extensive editing experiments across various volume datasets demonstrate that TexGS-VolVis outperforms existing NVS-based scene editing methods in terms of editing quality.
\vspace{-0.05in}
\end{myitemize}

\vspace{-0.05in}
\section{Related Work}

{\bf Image-based methods for VolVis.}
Image-based methods~\cite{IBM-survey} synthesize novel images for a 3D scene based on existing rendering results. 
These methods can be categorized into {\em image-based rendering} and {\em image-based modeling}.
In VolVis, \hot{image-based rendering} generates new volume rendering results directly from input images, bypassing the need for geometric modeling. 
Early works by Tikhonova et al.\ \cite{Tikhonova-CGF10, Tikhonova-PVIS10, Tikhonova-TVCG10} designed explorable images with multiple layers, enabling image-space color and opacity editing in post-hoc analysis. 
Frey et al.\ \cite{Frey-VDI} introduced {\em volumetric depth images}, a compact representation of volume data that facilitates accurate rendering near the views from which the \hot{volumetric depth images} were generated. 
Gupta et al.\ \cite{Gupta-ESPGV23} later proposed an efficient parallel algorithm for generating \hot{volumetric depth images} over distributed data.

Beyond data generation~\cite{Gu-CGA21, Han-CG22, Han-VI22, Yao-CG23, Gu-PVIS22, Tang-CG24} and neural compression~\cite{Tang-PVIS24, Gu-CG23, YF-Lu-VISSP24, Yang-PVISVN25, Son-VISSP25, Han-VIS25}, recent advancements have integrated deep learning with image-based rendering to enhance visualization generation~\cite{Wang-TVCG23}. 
For example, Berger et al.\ \cite{Berger-TVCG18} optimized a GAN to generate VolVis images from TFs and viewing parameters. 
Hong et al.\ \cite{Hong-PVIS19} introduced DNN-VolVis, a generative framework that synthesizes rendering images with the desired effects based on reference DVR images. 
Han and Wang~\cite{Han-TVCG23} developed CoordNet, a coordinate-based fully connected network for NVS using a set of multi-view visualization images. 
Additionally, surrogate models such as InSituNet~\cite{He-InsituNet}, VDLSurrogate~\cite{Shi-VDLSurrogate}, and ParamsDrag~\cite{Li-VIS24} leverage pretrained deep networks on large-scale DVR datasets rendered under different simulation and visualization parameters to preview rendering results for ensemble datasets.

Unlike \hot{image-based rendering}, \hot{image-based modeling} synthesizes new visualizations through a geometric modeling process. 
For instance, Niedermayr et al.\ \cite{Niedermayr-arxiv24} employed 3DGS~\cite{Kerbl-TOG23} for cinematic anatomy, enabling photorealistic rendering of medical data on consumer-grade devices. 
Tang et al.\ \cite{Tang-PVIS25} developed iVR-GS, utilizing editable Gaussian primitives to support real-time adjustments of color, opacity, and lighting for photorealistic rendering. 
Yao et al.\ \cite{Yao-PVIS25} introduced ViSNeRF, a multidimensional radiance field for NVS in dynamic VolVis scenes.
Yao and Wang~\cite{Yao-CG25} designed ReVolVE that reconstructs volumes from multi-view rendering images for visualization enhancement. 

Our TexGS-VolVis falls into the \hot{image-based modeling} category. 
Unlike the above methods, which focus on accurate reconstruction, it prioritizes expressive VolVis scene representation, enabling effective style transfer based on user-specified image or textual prompts. 
A recent work closely related to ours is StyleRF-VolVis~\cite{Tang-VIS24}, a NeRF-based solution that supports style transfer using reference style images. 
TexGS-VolVis surpasses StyleRF-VolVis by offering text-driven style transfer, real-time rendering, more realistic relighting with directional adjustments, and a composable model for scalable scene representation.

{\bf Differentiable primitive-based rendering.}
Primitive-based rendering techniques~\cite{Zwicker-VIS01, Zwicker-SIGGRAPH01, Sainz-CG04} generate images by rasterizing geometric primitives and have been widely studied for their computational efficiency. 
3DGS~\cite{Kerbl-TOG23} utilizes anisotropic Gaussian primitives to model a scene by optimizing multi-view images, delivering superior rendering speed and reconstruction quality compared to NeRF-like methods.

This foundational work has inspired a series of follow-ups to enhance the scene representation of 3DGS. 
For example, Gao et al.\ \cite{Gao-ECCV24} introduced relightable 3DGS, which extends the original 3D Gaussian primitives with {\em bidirectional reflectance distribution function} attributes, enabling realistic relighting of real-world scenes. 
Wu et al.\ \cite{Wu-arXiv24} developed 3DGUT that allows for rendering with distorted cameras using the unscented transform. 
To improve the geometry reconstruction accuracy of GS, Huang et al.\ \cite{Huang-SIGGRAPH24} proposed 2DGS, which utilizes 2D Gaussian primitives for perspective-correct splatting, as opposed to the original 3D primitives.
Building on this approach, several concurrent works—including GSTex~\cite{Rong-arXiv24}, BBSplat~\cite{Svitov-arXiv24}, and HDGS~\cite{Song-arXiv24}—have focused on improving the reconstruction accuracy of 2DGS by integrating the texture attribute into each 2D Gaussian primitive.

In this paper, TexGS-VolVis also adopts textured 2DGS for scene reconstruction. 
However, unlike~\cite{Rong-arXiv24, Svitov-arXiv24, Song-arXiv24}, our primary goal is not to enhance reconstruction accuracy. 
Instead, we leverage the capability of textured 2DGS to decouple the geometry and appearance representations of Gaussian primitives. 
This enables {\em geometry-consistent} editing while improving appearance expressiveness. 

{\bf Style transfer for 3D scenes.}
A central issue of NPR is transferring or mimicking a particular style while rendering a 3D scene. 
In visualization, previous studies have attempted to harness NPR techniques to create visually compelling results that are easier to interpret. 
For instance, Laidlaw~\cite{Laidlaw-CGF01} explored applying loose textures to fluid flow visualizations, mimicking the brush strokes of Van Gogh's oil paintings. 
Lu et al.\ \cite{Lu-VIS02} developed an NPR method to approximate the stipple style during volume rendering. 
Bruckner and Gr{\"o}ller~\cite{Bruckner-CGF07} introduced style TFs that integrate multiple NPR effects into a unified framework using lit spheres. 
Despite their effectiveness, traditional NPR methods often rely on manual rule design to emulate specific artistic styles.

To overcome this limitation, recent works have leveraged deep neural networks (e.g., VGG~\cite{Karen-VGG}) to enable style transfer for 3D scenes by editing the appearance parameters of 3DGS or NeRF models. 
For example, Zhang et al.\ \cite{Zhang-ARF} proposed a {\em nearest neighbor feature matching} (NNFM) loss function to transfer the artistic style from a reference image to a 3D scene represented by a NeRF model. 
Jain et al.\ \cite{Jain-arXiv24} introduced StyleSplat, which incorporates the NNFM loss into the 3DGS framework for faster style transfer. 
Haque et al.\ \cite{Haque-ICCV23} developed Instruct-NeRF2NeRF (IN2N), enabling stylization of NeRF scenes using text instructions by iteratively editing input images with InstructPix2Pix (IP2P)~\cite{brooks-InstructPix2Pix}. 
Chen et al.\ \cite{Chen-DGE} proposed DGE, which improves consistency in text-driven scene editing by incorporating epipolar constraints into the diffusion model. 
Distinct from these methods, our TexGS-VolVis represents the scene using textured 2D Gaussian primitives, resulting in significantly faster rendering speeds than NeRF-based methods while offering greater editing flexibility.

{\bf Large pretrained models.} 
Recent advances in large pretrained models~\cite{Karen-VGG, Radford-CLIP, Rombach-SD, Kirillov-ICCV23} have empowered researchers to tackle various challenging downstream tasks in computer vision. 
For instance, Patashnik et al.\ \cite{Patashnik-ICCV21} introduced StyleCLIP, which leverages the pretrained CLIP~\cite{Radford-CLIP} model to enable text-based interaction for manipulating images generated by StyleGAN~\cite{Karras-TPAMI21}. 
Ma et al.\ \cite{Ma-NC24} developed MedSAM, achieving state-of-the-art segmentation performance on medical data by fine-tuning the {\em segment anything model} (SAM)~\cite{Kirillov-ICCV23} with over one million medical image-mask pairs. 
Ye et al.\ \cite{Ye-TOG24} proposed StableNormal, which adapts the diffusion priors of the pretrained stable diffusion~\cite{Rombach-SD} model to perform accurate monocular normal estimation. 
Unlike these approaches that rely on a single large pretrained model, this paper harnesses the capabilities of multiple pretrained models, including SAM~\cite{Kirillov-ICCV23}, CLIP~\cite{Radford-CLIP}, and IP2P~\cite{brooks-InstructPix2Pix}, to effectively segment and stylize VolVis scenes.

\begin{figure*}[htb]
\centering
\includegraphics[width=1\linewidth]{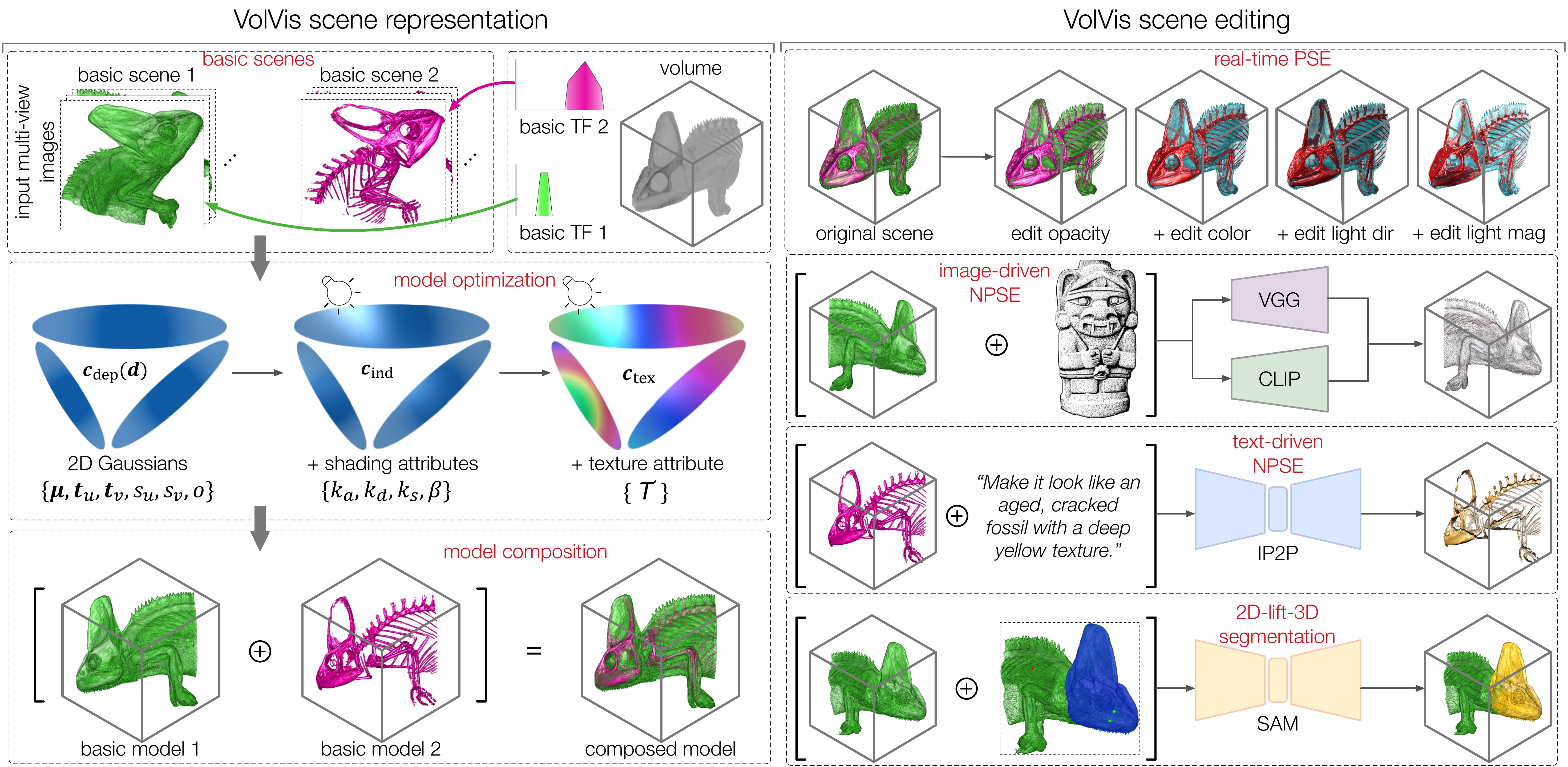}
\vspace{-.2in}
\caption{The workflow of TexGS-VolVis consists of two stages: scene representation (left) and scene editing (right). In the scene representation stage, each basic scene is trained using multi-view images and 2D Gaussian primitives enhanced with shading and texture attributes. The composed model is generated by combining the basic models without requiring additional optimization. In the scene editing stage, users can perform real-time PSE, image- and text-driven NPSE, and 2D-lift-3D segmentation for partial editing.}
\label{fig:workflow}
\end{figure*}

\vspace{-0.05in}
\section{TexGS-VolVis}
\label{sec:method}

Figure~\ref{fig:workflow} illustrates the two-stage TexGS-VolVis workflow: scene {\em representation} and {\em editing}. 
\hot{To avoid restricting model visibility to a single preset TF, we train multiple basic models, each on a basic scene defined by a disjoint basic TF.
The training process for each basic model consists of three phases. 
First, we train a vanilla 2DGS (Section~\ref{subsec:vanilla2DGS}) to initialize the geometry representation. 
Next, we augment each primitive with shading attributes and compute the view-dependent color using the Blinn-Phong shading model (Section~\ref{subsec:BlinnPhongShadingModel}), producing a relightable 2DGS (Section~\ref{subsec:Relightable2DGS}). 
Finally, we further enhance expressiveness by adding and optimizing a texture attribute for each primitive (Section~\ref{subsec:TexturedGaussianPrimitives}).
Training a basic model typically takes several minutes, with the second phase—optimizing shading attributes—accounting for approximately 70\% of the total time.}

\hot{Once TexGS-VolVis is optimized for scene representation, it supports a range of both online and offline editing scenarios.
For online editing, users can perform real-time PSE to interactively modify scene attributes (color, opacity, and lighting) without additional training (Section~\ref{subsec:RecolorableTextureAndPSE}).
For offline editing, more advanced capabilities are available, including image- and text-driven NPSE (Section~\ref{subsec:ImageTextNPSE}) and {\em 2D-lift-3D scene segmentation} (Section~\ref{subsec:2D-lift-3DSeg}).
These require additional optimization but enable more expressive and flexible scene manipulation. 
Image-driven NPSE typically takes around ten minutes, due to frequent invocations of large pretrained models (VGG and CLIP) during the offline optimization process. 
In contrast, text-driven NPSE and 2D-lift-3D scene segmentation usually complete within one to three minutes, as they rely on fewer calls to large pretrained models (IP2P and SAM).
}

\vspace{-0.05in}
\subsection{\hot{Preliminaries}}

\hot{
\subsubsection{Vanilla 2DGS}
\label{subsec:vanilla2DGS}
}

Unlike existing NVS methods for VolVis that support real-time rendering~\cite{Niedermayr-arxiv24, Tang-PVIS25}, we opt for 2DGS~\cite{Huang-SIGGRAPH24} over 3DGS~\cite{Kerbl-TOG23} to represent the scene.
In 2DGS, each Gaussian primitive is modeled as a 2D elliptical surfel, rather than a 3D ellipsoid as in 3DGS. 
Each surfel is parameterized by 
view-dependent color $\mathbf{c}_{\dep}$, represented using optimizable {\em spherical harmonic} coefficients,
opacity $o$,
mean position $\bm{\mu}$,
two principal tangential axis vectors $\mathbf{t}_u$ and $\mathbf{t}_v$, paired with scaling factors $s_u$ and $s_v$.
The surfel normal $\mathbf{n}$ is then computed as $\pm(\mathbf{t}_u \times \mathbf{t}_v)$, where the sign depends on the viewing direction.

Given a pixel location $\mathbf{x}$ in image space, instead of rasterizing the primitive center onto the image plane and approximating the projected covariance as in 3DGS, 2DGS directly computes the ray-splat intersection $(u, v)$ within the surfel coordinate system. 
This is accomplished using the projection matrix $\mathbf{P}$ and the splat-to-world transformation $\mathbf{H}$, as follows 
\begin{equation}
\mathbf{x}=\mathbf{P}\mathbf{H}(u,v,1,1)^{T},
\end{equation}
where $\mathbf{H}$ is parameterized by the primitive properties as
\begin{equation}
	\mathbf{H}=\begin{bmatrix}
s_u\mathbf{t}_u & s_u\mathbf{t}_v & 0 & \bm{\mu} \\
0 & 0 & 0 & 1
\end{bmatrix},
\end{equation}
The final pixel color $\mathbf{C}$ is computed by accumulating contributions from all $N$ intersected surfels along the ray using alpha blending
\begin{equation}
	\mathbf{C}=\sum_{i\in N}T_io_iG_i(\mathbf{x})(\mathbf{c}_{\dep})_i(\mathbf{d}) \quad \textrm{with} \quad T_i=\prod^{i-1}_{j=1}(1-o_jG_j(\mathbf{x})),
\label{eqn:alpha-blending}
\end{equation}
where $T$ represents the accumulated transmittance, $G_i(\mathbf{x})$ is the weight of the $i$-th surfel, and $\mathbf{d}$ denotes the viewing direction.
In practice, depth and normal maps for 2DGS can be obtained similarly to Equation~\ref{eqn:alpha-blending}. 

\hot{
\vspace{-0.05in}
\subsubsection{Blinn-Phong Shading Model}
\label{subsec:BlinnPhongShadingModel}

The Blinn-Phong shading model is widely used for rendering VolVis scenes with lighting effects.
Given a light direction $\mathbf{l}$, the view-dependent color $\mathbf{c}_{\dep}$ emitted by a sample voxel toward the camera is computed as the sum of ambient ($\mathbf{c}_a$), diffuse ($\mathbf{c}_d$), and specular ($\mathbf{c}_s$) components 
\begin{subequations}
\label{eqn:Blinn-Phong}
\begin{align}
	\mathbf{c}_a&=k_a\mathbf{I}_a, \\
	\mathbf{c}_d&=k_d\mathbf{I}_d |\mathbf{n} \cdot \mathbf{l}|,\\
	\mathbf{c}_s&=
	\begin{cases}
		k_s\mathbf{I}_s |\mathbf{n} \cdot \mathbf{h}|^{\beta},	& \text{if } |\mathbf{n} \cdot \mathbf{l}| > 0 \\
		0,		& \text{otherwise} 
	\end{cases}
\end{align}
\end{subequations}
Here, $(k_a, \mathbf{I}_a)$, $(k_d, \mathbf{I}_d)$, and $(k_s, \mathbf{I}_s)$ denote the ambient, diffuse, and specular coefficients and their corresponding material's color properties.
In practice, both $\mathbf{I}_a$ and $\mathbf{I}_d$ are typically set to the voxel color obtained from the TF based on voxel intensity, while $\mathbf{I}_s$ is usually set to white.
The shininess factor $\beta$ controls the sharpness of specular highlights. 
The surface normal $\mathbf{n}$ is derived from the gradient at the voxel position, and 
$\mathbf{h} = \frac{\mathbf{v} + \mathbf{l}}{|\mathbf{v} + \mathbf{l}|}$ is the halfway vector between the viewing direction $\mathbf{v}$ and the light direction $\mathbf{l}$.
}

\vspace{-0.05in}
\subsection{Relightable 2DGS}
\label{subsec:Relightable2DGS}

Our pipeline begins by optimizing the relightable 2DGS to accurately reconstruct the scene geometry while separating the lighting component from the appearance of the VolVis scene.
This optimization process consists of two main steps:
(1) obtain an initial representation using vanilla 2DGS and 
(2) augment each 2D Gaussian primitive with learnable Blinn-Phong shading attributes to enable lighting decomposition and relighting.
\hot{In the first step, we represent the view-dependent color $\mathbf{c}_{\dep}$ using spherical harmonic coefficients.
In the second step, rather than relying on these coefficients to compute $\mathbf{c}_{\dep}$ for each primitive, we adopt the Blinn-Phong shading model (Equation~\ref{eqn:Blinn-Phong}) for this purpose.
Specifically, we augment each initialized Gaussian primitive with additional shading attributes, including the view-independent color $\mathbf{c}_{\ind}$, ambient, diffuse, and specular coefficients ${k_a, k_d, k_s}$, and the shininess factor $\beta$.
We replace the TF-sampled voxel color in Equation~\ref{eqn:Blinn-Phong} with the learnable $\mathbf{c}_{\mathrm{ind}}$, which serves as both $\mathbf{I}_a$ and $\mathbf{I}_d$, while $\mathbf{I}_s$ is set to white.
The normal vector $\mathbf{n}$ for each Gaussian is computed from its $\mathbf{t}_u$ and $\mathbf{t}_v$ attributes (see Section~\ref{subsec:vanilla2DGS}), and the halfway vector $\mathbf{h}$ is derived from the viewing and light directions provided in the training data.
}

\hot{To optimize the vanilla 2DGS}, we randomly initialize Gaussian primitives and optimize them using the reconstruction loss $L_r$ for 10,000 iterations, defined as
\begin{equation}
	L_{r}=L_{c}+\lambda_{n} L_n + \lambda_{\alpha} L_\alpha,
	\label{eqn:loss}
\end{equation}
where $L_{c}$ and $L_\alpha$ are pixel-wise reconstruction losses for the RGB and alpha channels, respectively, combining L1 loss with the D-SSIM term~\cite{Niedermayr-arxiv24, Tang-PVIS25}. 
$L_{n}$ represents the {\em normal consistency regularization}~\cite{Huang-SIGGRAPH24}. 
$\lambda_{n}$ and $\lambda_{\alpha}$ are set to 0.05 and 0.1, respectively. 
We do not incorporate {\em depth distortion regularization}~\cite{Huang-SIGGRAPH24} in the standard 2DGS optimization.
Empirically, while depth regularization can slightly improve geometry representation, it sacrifices NVS reconstruction accuracy on VolVis scene datasets, offering minimal practical benefit. 
Upon completing this optimization, we establish a solid initial representation of the scene's geometry and appearance.

\hot{When optimizing the shading attributes, we first initialize $\mathbf{c}_{\ind}$ using the zeroth-order spherical harmonic coefficient trained in the first step to accelerate convergence.
We then optimize all primitive attributes for an additional 20,000 iterations.}
During optimization, in addition to $L_{r}$, we apply {\em bilateral smoothness regularization}~\cite{Gao-ECCV24, Tang-PVIS25} to prevent excessive variation in shading attributes within smooth-color regions.
Let $\mathbf{c}_{\gt}$ be the ground-truth pixel color from multi-view images, \hot{$({k_s})_i$} is the specular coefficient attribute of the $i$-th Gaussian. 
We define the rendered specular coefficient map as $K_{s}=\sum_{i\in N}T_io_iG_i(\mathbf{x})(k_s)_i$.
The bilateral smoothness regularization on the specular attribute is then formulated as
\begin{equation}
(L_b)_s=|\bigtriangledown K_{s}|\exp{(-|\bigtriangledown\mathbf{c}_{\gt}|)},
\end{equation}
which encourages smooth transitions in specular coefficients, weighted by image gradients to preserve high-frequency details. 
We apply the same regularization to other shading attributes, including $k_a$, $k_d$, and $\beta$, to obtain $(L_b)_a$, $(L_b)_d$, and $(L_b)_{\beta}$. 
To balance these terms with $L_{r}$, all \hot{regularization terms} are scaled by a factor of 0.01 during joint Gaussian optimization.
Once training is complete, the additional shading attributes effectively separate the lighting component from the scene's appearance. 
Given an unseen light source and camera position, relightable 2DGS can synthesize novel lighting effects following the Blinn-Phong shading model. 
In the subsequent editing stage, the geometry and shading attributes of each primitive are frozen, preserving accurate geometry and lighting information for the VolVis scene.

\vspace{-0.05in}
\subsection{Textured Gaussian Primitives}
\label{subsec:TexturedGaussianPrimitives}

In conventional GS~\cite{Kerbl-TOG23, Huang-SIGGRAPH24, Tang-PVIS25}, each Gaussian primitive can only represent a single color and shape for a given camera view.
Since each primitive encodes appearance and geometry, these attributes are tightly coupled. 
When editing the appearance of a VolVis scene while preserving its content geometry, this limited expressive capacity significantly hinders the ability to capture complex editing effects (e.g., image- or text-driven NPSE). 
To overcome this limitation, we propose assigning a texture map to each relightable 2D Gaussian primitive, enabling even a single Gaussian to capture intricate appearance details.
By leveraging {\em per-primitive texturing}, our method decouples appearance representation from the scene's geometric topology and complexity, ensuring geometry-consistent NPSE. 
We further analyze the rationale behind this design choice in Section~\ref{subsec:ablation}.

For the $i$-th relightable 2D Gaussian optimized in the first stage, we assign a fixed-resolution 2D texture map $\altmathcal{T}_i$ of size $U_i \times V_i \times 3$, where each texel corresponds to an RGB value.
The UV mapping for each 2D Gaussian is constructed so that each texel represents a square region in world space with a uniform size of $T_{\size} \times T_{\size}$ across all Gaussians.
To ensure adequate coverage of the high-weight regions in $G_i(\mathbf{x})$ (Equation~\ref{eqn:alpha-blending}), the texture dimensions $U_i$ and $V_i$ are determined based on the scaling factors $(s_u)_i$ and $(s_v)_i$ of the $i$-th Gaussian.
Specifically, $U_i=\lceil 6 \times (s_u)_i /T_{\size} \rceil$ and \hot{$V_i=\lceil 6 \times (s_v)_i /T_{\size} \rceil$}. 
This adaptive allocation ensures that larger-sized primitives receive more texels and smaller-sized primitives are assigned fewer texels, facilitating a level-of-detail adjustment based on the spatial extent of the Gaussians. 
The texel dimension $T_{\size}$ is computed as $T_{\size}=T_{\total} / \sum_{i \in N}(36 \times (s_u)_i (s_v)_i)$, where $T_{\total}$ is a hyperparameter that controls the total number of texels allocated across all Gaussian.

When rendering textured Gaussians, a ray is cast from the camera origin, resulting in an intersection point $\mathbf{p}_i$ with the $i$-th Gaussian.
The corresponding UV coordinate $(u_i, v_i)$ on the texture map $\altmathcal{T}_i$ for $\mathbf{p}_i$ is computed as
\begin{subequations}
\begin{align}
	u_i (\mathbf{p}_i)&=((\mathbf{t}_u)_i \cdot \mathbf{p}_i)/T_{\size}+(U_i - 1)/2, \\
	v_i (\mathbf{p}_i)&=((\mathbf{t}_v)_i \cdot \mathbf{p}_i)/T_{\size}+(V_i - 1)/2.
\end{align}
\end{subequations}
Finally, the color $(\mathbf{c}_{\tex})_i$ of the $i$-th Gaussian at $\mathbf{p}_i$ is obtained by bilinear interpolation on the texture map using $u_i$ and $v_i$.
During rendering, the sampled texture color $\mathbf{c}_{\tex}$ allows each Gaussian primitive to exhibit spatially varying colors at different ray-intersection points rather than a single uniform color.

\vspace{-0.05in}
\subsection{Recolorable Textures and PSE}
\label{subsec:RecolorableTextureAndPSE}

Although textured Gaussian primitives can capture more detailed appearances using spatially varying colors, the colors of the represented VolVis scene remain fixed during inference, limiting user exploration. 
To address this limitation, we extend textured Gaussian primitives with {\em recolorable textures}. 
Since each basic scene typically contains a dominant color, we represent the view-independent color $\mathbf{c}_{\ind}$ of the $i$-th Gaussian at the intersection point $\mathbf{p}_i$ as $\mathbf{c}_{\palette} + (\mathbf{c}_{\tex})_i(\mathbf{p}_i)$, where $\mathbf{c}_{\palette}$ is a learnable palette color parameter. 
This palette color is initialized as the mean color of input multi-view images and is shared across all Gaussians within the basic scene.
We optimize both the recolorable texture and $\mathbf{c}_{\palette}$ for all Gaussians using an RGB reconstruction loss $L_c$ (refer to Equation~\ref{eqn:loss}) combined with a {\em sparsity regularization} term~\cite{Tang-PVIS25, Tang-VIS24}. 
The sparsity regularization is implemented as an L1 loss on the rendered color map of $\mathbf{c}_{\tex}$, encouraging the sampled $\mathbf{c}_{\tex}$ to remain minimal and preventing significant shifts in $\mathbf{c}_{\ind}$ from $\mathbf{c}_{\palette}$.

After optimizing the recolorable texture, TexGS-VolVis enables real-time scene editing, supporting PSE, including color, opacity, and lighting adjustments during inference.
For each basic scene, color editing is achieved by adjusting $\mathbf{c}_{\palette}$.
Opacity is controlled by scaling the Gaussian opacity attributes with a user-specified factor. 
Lighting adjustments are made by scaling the parameters $k_a$, $k_d$, $k_s$, and $\beta$. 
Since TexGS-VolVis uses Blinn-Phong shading for rendering, the scene can be relit in real time when users modify the lighting direction.

\vspace{-0.05in}
\subsection{Image- and Text-Driven NPSE}
\label{subsec:ImageTextNPSE}

TexGS-VolVis leverages {\em per-primitive texturing} to decouple appearance representation from geometry modeling, enabling geometry-consistent NPSE and greatly enhancing the expressive capacity of Gaussian primitives to capture complex style patterns. 
By adopting TexGS-VolVis as a flexible and editable VolVis scene representation method, we propose three approaches incorporating various large pretrained models to facilitate different scene editing tasks, as described below.

{\bf Image-driven NPSE.}
Like StyleRF-VolVis~\cite{Tang-VIS24}, TexGS-VolVis supports NPSE based on an image prompt containing a reference style pattern. 
If users are particularly interested in the style of a local region within the reference image, a pretrained 2D segmentation model, such as SAM~\cite{Kirillov-ICCV23}, can be used to extract the desired style regions, following the same approach as in StyleRF-VolVis. 
During image-driven NPSE, we randomly initialize all Gaussian texture maps to represent the stylized appearance and then use the texture-sampled color to define the view-independent color without involving the palette color. 
All other attributes optimized in the relightable 2DGS stage, including geometry attributes (e.g., $\bm{\mu}$) and shading attributes (e.g., $\beta$), are fixed to preserve the consistent geometry and lighting of the VolVis scene's content.
Given a pre-trained CLIP model $\altmathcal{C}$, an image prompt $\altmathcal{I}_{\prompt}$, and an image $R$ rendered from TexGS-VolVis under a randomly selected training view with Blinn-Phong shading applied, 
the loss $L_{\img}$ for image-driven NPSE is formulated as
\begin{equation}
	L_{\img} = \lambda_s L_{\nnfm}(R,\altmathcal{I}_{\prompt}) + (1 - \lambda_s)||\altmathcal{C}(R)-\altmathcal{C}(\altmathcal{I}_{\prompt})||_2,
\end{equation}
where the first term is the NNFM loss, commonly used in NeRF-based stylizations~\cite{Zhang-ARF, Tang-VIS24}, which employs a pretrained VGG-16 network to capture fine stylization details.
The limitation of the NNFM loss is that it may lead to a stylized appearance that emphasizes a local pattern from $\altmathcal{I}_{\prompt}$, potentially diverging from the overall style.
To mitigate this, we incorporate a CLIP-based loss as the second term, which encourages TexGS-VolVis to match the global style pattern in $\altmathcal{I}_{\prompt}$ by calculating the L2 distance between feature vectors extracted by the CLIP model $\altmathcal{C}$.
The hyperparameter $\lambda_s$ is used to balance between fine stylization details and global style consistency.
In this paper, we set $\lambda_s = 0.9$ and optimize TexGS-VolVis for 3,000 iterations to perform image-driven NPSE unless otherwise specified.

{\bf Text-driven NPSE.}
Building upon existing NeRF and 3DGS scene editing approaches~\cite{Haque-ICCV23, Dong-NeurIPS23, Chen-CVPR24, Chen-DGE} guided by text instructions, we propose a novel text-driven NPSE method specifically designed for VolVis scenes represented by TexGS-VolVis. 
Given an NVS model (such as NeRF or 3DGS) and the multi-view images used to reconstruct it, the mainstream 3D text-driven editing approach utilizes the pretrained IP2P model~\cite{brooks-InstructPix2Pix}.
This model takes unedited multi-view images and a text instruction as prompts to generate edited images, which are then used to fine-tune the NVS model.
However, due to inconsistent editing effects across the edited multi-view images, existing methods typically rely on an {\em iterative updating strategy}. 
This strategy gradually updates the reconstructed scene by iteratively editing the multi-view images while optimizing the NVS model. 
While this approach enables multi-view and geometry-consistent editing in simple, opaque real-world scenes, it faces challenges when applied to complex and semi-transparent VolVis scenes, often resulting in blurry appearances or geometry-inconsistent scene edits.

To address these challenges, we leverage TexGS-VolVis to represent the VolVis scene and use texture-sampled color to represent the view-independent color, as in image-driven NPSE. 
We also introduce three novel strategies compared to previous editing frameworks:
\begin{myitemize}
\vspace{-0.05in}
\item {\bf Geometry-consistent editing}:\ Since TexGS-VolVis decouples appearance and geometry through the texture attribute, geometry-consistent editing can be achieved by optimizing only the texture attribute with an RGB reconstruction loss. 
This eliminates the need for iterative updates to maintain geometry consistency. 
Instead, we adopt a {\em one-time updating strategy}, where texture parameters are updated based on a single round of IP2P editing for the editing views. 
In our setup, we update TexGS-VolVis for 1,500 iterations using the IP2P editing images. 
\item {\bf Reduced training views}:\ Since geometry consistency is no longer a concern, we can update TexGS-VolVis using only a small subset of training views—specifically six views that cover the scene's front, back, left, right, top, and bottom. 
This reduces the ambiguity caused by inconsistent editing effects across different views of the same scene region, resulting in stylization outcomes with clear boundaries. 
Further ambiguity reduction can be achieved by incorporating the epipolar constraint~\cite{Chen-DGE} into the IP2P model's inference process.
\item {\bf Opacity adjustment}:\ We perform PSE to adjust the opacity of the TexGS-VolVis scene during optimization, enhancing editing results. 
This adjustment stems from our observation that the opacity of the VolVis scene directly impacts text-driven NPSE. 
When the opacity is low, each editing view can influence the appearance of all Gaussian primitives within the scene, leading to blurry stylization due to inconsistent IP2P edits. 
Conversely, high opacity can cause occlusion, leaving uncovered Gaussian primitives unedited. 
In practice, we initialize the scene's opacity at 2$\times$ its target value and gradually reduce it to 0.5$\times$ over fixed iteration intervals.
\vspace{-0.05in}
\end{myitemize}

\subsection{2D-lift-3D Scene Segmentation}
\label{subsec:2D-lift-3DSeg}

To apply different editing effects to distinct regions of the same basic scene, a {\em 2D-lift-3D scene segmentation}~\cite{Cen-arXiv23, Liu-CVPR24, Choi-ECCV24, Hu-arXiv24} approach is required. 
Specifically, we first use SAM to generate a 2D mask based on user-specified point prompts on an image rendered by TexGS-VolVis from any arbitrary view. 
This 2D segmentation is lifted into 3D by identifying and segmenting the corresponding Gaussian primitives. 
This process results in distinct Gaussian groups within the same basic scene, enabling 3D segmentation of the scene represented by Gaussians. 
Finally, PSE or NPSE techniques can be applied to various groups of Gaussian primitives, allowing for more flexible and controllable VolVis scene editing.

To achieve 2D-lift-3D segmentation of a scene represented by Gaussian primitives, recent works~\cite{Cen-arXiv23, Choi-ECCV24} have tackled this problem by integrating and optimizing a learnable feature vector within each Gaussian. 
While effective, the additional training time and storage costs associated with these feature vectors are prohibitive in our case, where multiple basic scenes, each potentially containing millions of Gaussians, construct a single VolVis scene. 
Alternatively, we adopt a {\em training-free segmentation strategy} inspired by SAGD~\cite{Hu-arXiv24}. 
Given a set of user-specified point prompts on a reference view, we use SAM to automatically generate a series of 2D segmentation masks for the corresponding parts across all training views. 
We then perform {\em label propagation} to assign distinct labels to each 2D Gaussian and apply the {\em multi-view label voting} technique~\cite{Hu-arXiv24} to determine the segmented Gaussian primitives.

\begin{table}[htb]
\caption{The datasets and their settings for image-driven and text-driven NPSE. The rendering time is for DVR using ParaView.}
\vspace{-0.1in}
\centering
\resizebox{\columnwidth}{!}{
\begin{tabular}{c|cc|ccc}
   	     & volume     &image  		&\# basic &  volume  & rendering     \\ 
 dataset & resolution &resolution   &  scenes &  size (MB)  &  time (ms) \\ \hline
 five jets &256$\times$256$\times$256 &800$\times$800 &2 &64  &29.68 \\
 mantle & 360$\times$201$\times$180 &800$\times$800 &2 &49.7  &20.49  \\
 supernova &864$\times$864$\times$864 &800$\times$800 &2 &2,460.3  &132.33  \\
 wood &1024$\times$1024$\times$1024 &800$\times$800 &1 &4,096  &153.69  \\ \hline
  beetle &832$\times$832$\times$494 &512$\times$512 &1 &1,304.4 & 71.75  \\ 
 chameleon & 1024$\times$1024$\times$1080 &512$\times$512 &2 &4,320  &152.8  \\
   engine	   & 256$\times$256$\times$128 &512$\times$512 &2 &32 &16.14  \\
 ionization	   & 600$\times$248$\times$248 &512$\times$512 &2 &140.8 &30.42  \\ 
\end{tabular}
}
\label{tab:dataset}
\end{table}

\begin{figure*}[htb]
\begin{center}
	\includegraphics[width=0.975\linewidth]{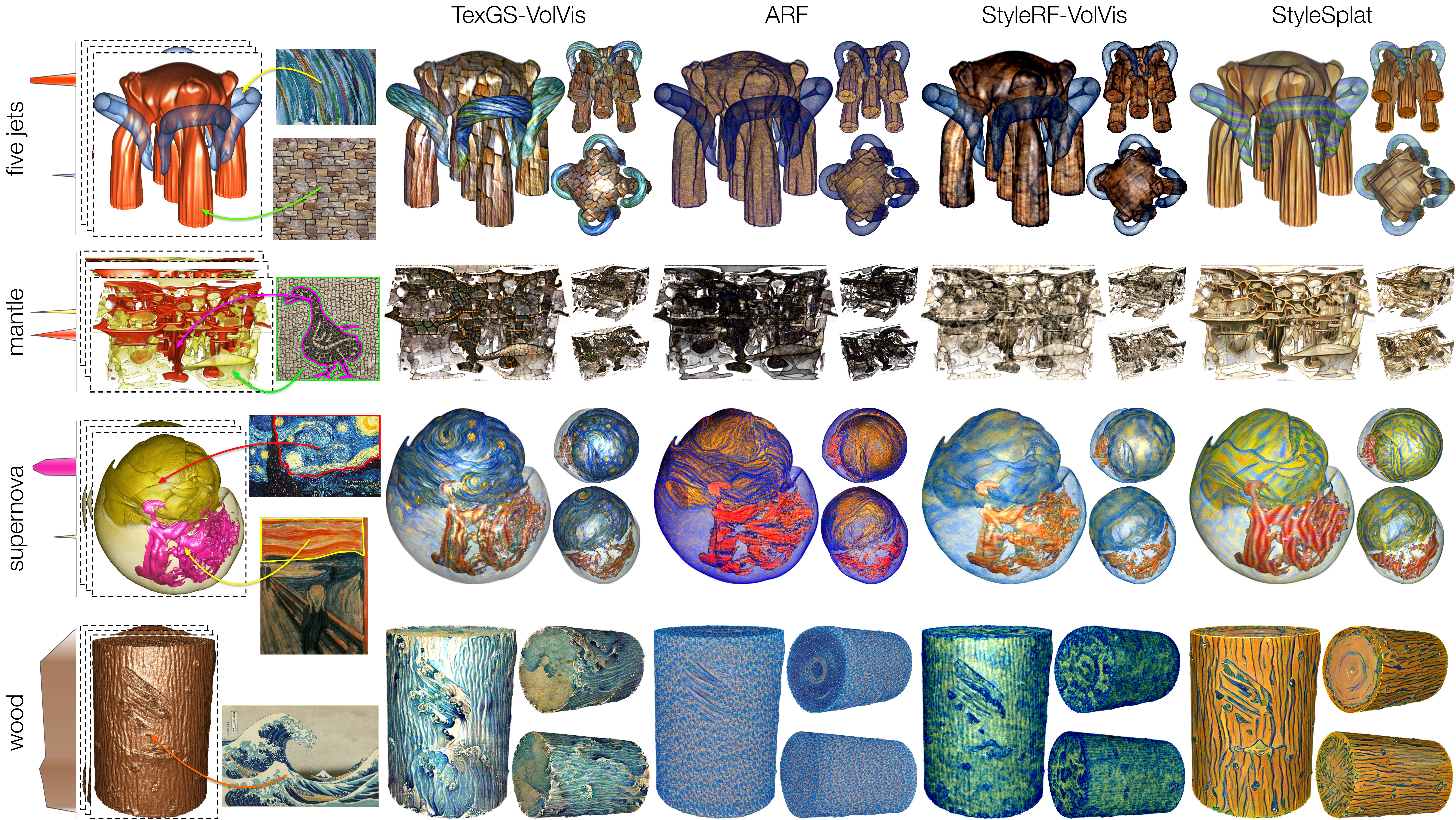}
	\end{center}
	\vspace{-.2in}
\caption{Comparison of image-driven NPSE results. 
\hot{The leftmost panel shows the multi-view images rendered using DVR, along with their corresponding TFs.
Each color in a TF represents a distinct basic scene.}
The arrows indicate the correspondence between styles and scenes. 
The five jets and wood datasets use the entire reference images, while the mantle and supernova datasets utilize selected style regions from the reference images. TexGS-VolVis retains the original scene's lighting more faithfully while enabling adjustments to light magnitude and direction during inference.} 
\label{fig:baseline-img-NPSE}
\end{figure*}

\vspace{-0.2in}
\section{Results and Discussion}

\subsection{Datasets and Training}

To demonstrate the superior editing quality and flexibility of TexGS-VolVis, we compare it with other state-of-the-art 3D scene style transfer methods using the datasets listed in Table~\ref{tab:dataset}. 
For each basic scene in a dataset, we render 162 multi-view images using icosphere sampling to optimize TexGS-VolVis for scene representation. 
All images are rendered with ParaView using the NVIDIA IndeX plugin to ensure appropriate lighting effects. 
For text-driven NPSE, we follow previous works~\cite{Haque-ICCV23, Dong-NeurIPS23, Chen-CVPR24, Chen-DGE} and set the image resolution to 512$\times$512 to avoid GPU out-of-memory errors when running the large pretrained model IP2P. 
As an \hot{image-based modeling} method, the rendering resolution of TexGS-VolVis is independent of the input multi-view image resolution. 
However, we maintain the same output image resolution as the input multi-view images to ensure a fair comparison of rendering speed between DVR and our method.

We implemented TexGS-VolVis using PyTorch and customized CUDA kernels for texture mapping and lighting computation. 
The Adam optimizer is used to optimize the attribute values of all Gaussian primitives. 
We set the learning rates as follows: 0.01 for additional shading attributes and $\mathbf{c}_{\palette}$, 0.025 for the texture attribute, and the same learning rate as 2DGS~\cite{Huang-SIGGRAPH24} for other attributes. 
For each basic scene, we set $T_{\total}$ to $1 \times 10^7$ and use the allocation method described in Section~\ref{subsec:TexturedGaussianPrimitives} to automatically assign an appropriate number of texels for each Gaussian primitive. 
All experiments, including DVR to generate multi-view images, were conducted on a local workstation with an NVIDIA RTX4090 GPU. 
When using the stable diffusion model IP2P for text-driven NPSE, we set its classifier-free guidance scales to 1.25 for the input multi-view images and 12.5 for text prompts.

\vspace{-0.05in}
\subsection{Image-Driven NPSE}

{\bf Baselines.}
We compare TexGS-VolVis against three baseline methods for image-guided NPSE:
\begin{myitemize}
\vspace{-0.05in}
\item ARF~\cite{Zhang-ARF} is a NeRF-based method that stylizes 3D scenes by training the Plenoxels model~\cite{plenoxels-CVPR22} and refining it with NNFM loss to ensure detailed style preservation and multi-view consistency.
\item StyleRF-VolVis~\cite{Tang-VIS24} is a NeRF-based method that uses a palette color network to classify different color regions and an unrestricted color network to represent stylized scene appearance. By leveraging the classification result and optimizing the unrestricted color network using NNFM, the framework enables diverse stylization effects across different color regions of the scene.
\item StyleSplat~\cite{Jain-arXiv24} is a 3DGS-based method that achieves stylization by freezing geometry attributes of Gaussian primitives and fine-tuning spherical harmonic coefficients of 3D Gaussian primitives using NNFM to align them with reference style images.
\vspace{-0.05in}
\end{myitemize}
\noindent Unlike other methods, StyleRF-VolVis utilizes implicit neural networks to represent 3D scenes. 
As a result, different basic models in StyleRF-VolVis cannot be directly combined to form a complete VolVis scene. 
Consequently, the input multi-view images for StyleRF-VolVis are rendered using a comprehensive TF, which is constructed by combining all the basic TFs and leveraging the palette color network to differentiate and stylize various color regions.

\begin{figure*}[htb]
\begin{center}
	\includegraphics[width=0.975\linewidth]{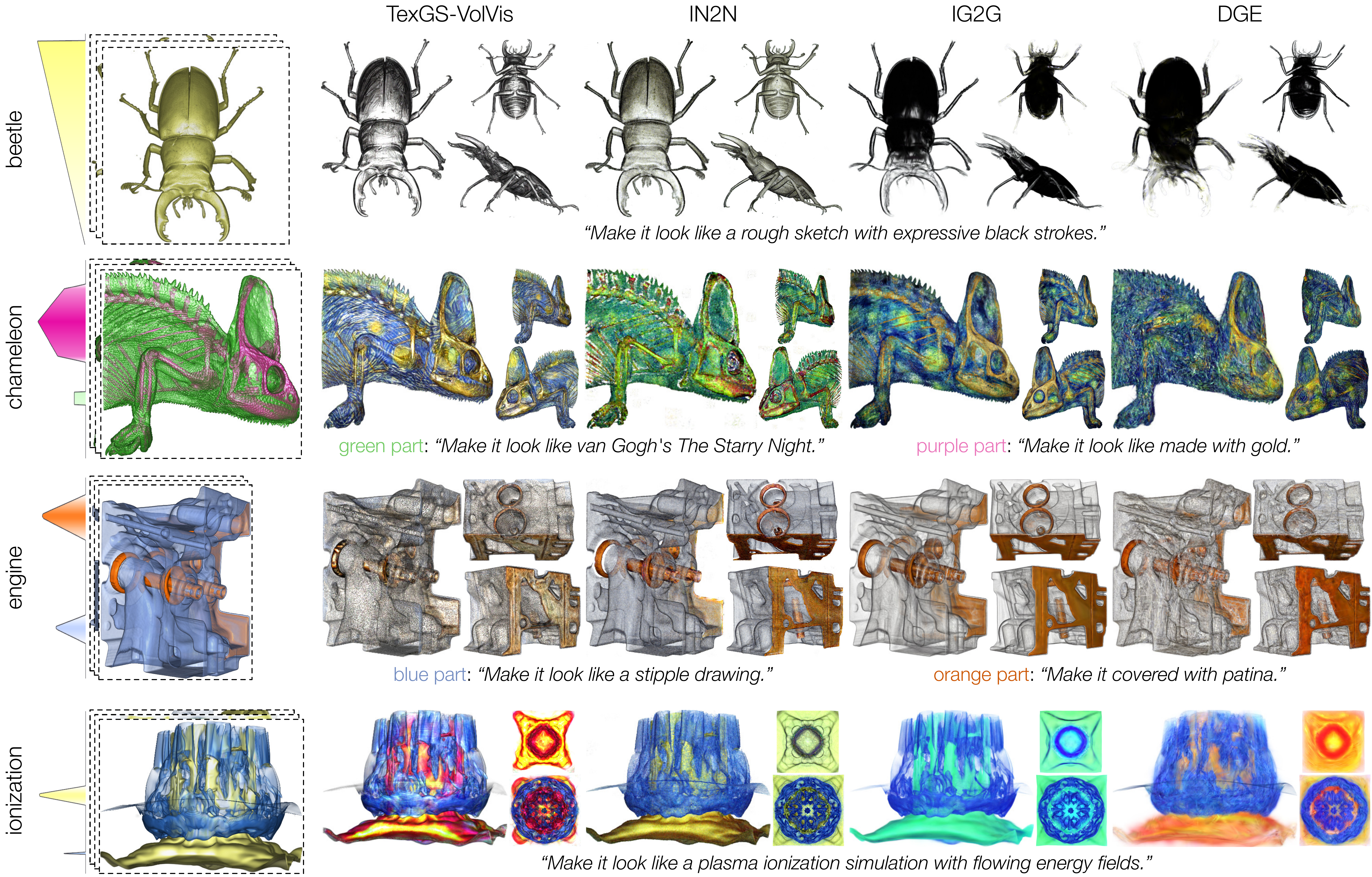}
	\end{center}
	\vspace{-.2in}
\caption{Comparison of text-driven NPSE results.
\hot{The leftmost panel shows the multi-view images rendered using DVR, along with their corresponding TFs.
Each color in a TF represents a distinct basic scene.}
The beetle and ionization datasets use a single prompt for the entire scene, while the chameleon and engine datasets employ two prompts for different scene parts specified by the TFs. 
TexGS-VolVis achieves superior lighting preservation and more expressive scene editing outcomes.} 
\label{fig:baseline-text-NPSE}
\end{figure*}

{\bf Qualitative evaluation.}
To demonstrate the superior image-driven NPSE quality of TexGS-VolVis, we compare it with baseline methods using VolVis scenes from the five jets, mantle, supernova, and wood datasets. 
Each scene is stylized by one or more styles extracted from the entire or a part of the reference style images. 
Figure~\ref{fig:baseline-img-NPSE} presents a qualitative comparison, showcasing rendering results from three different views for each stylization.
Among all methods, TexGS-VolVis most effectively preserves the original scene's lighting, delivering significantly clearer rendering results. 
During image-driven NPSE, ARF discards the view-direction input, representing the stylized scene with only view-independent colors. 
While this improves cross-view consistency, ARF fails to preserve the view-independent lighting of the original VolVis scene, resulting in blurry renderings that obscure intricate structures within complex volumes (e.g., mantle). 
For the other baseline methods, StyleRF-VolVis restores view-dependent lighting using an independent MLP network, and StyleSplat retains some lighting effects through its high-order \hot{spherical harmonic} coefficients. 
However, both methods can only preserve plausible lighting effects and cannot accurately relight the stylized scene regarding light magnitude and direction. 
For example, the rendering results of StyleRF-VolVis and StyleSplat on the five jets and wood datasets show significantly weaker lighting than the original VolVis scene illumination.
In contrast, TexGS-VolVis integrates the Blinn-Phong shading model within its inference process, preserving lighting effects and allowing for faithful lighting restoration. 
This enables interactive light direction and magnitude adjustments during rendering. 

In terms of stylization quality, TexGS-VolVis produces more detailed and expressive results compared to other methods. 
This superior quality can be attributed to the representation of the texture attribute within the Gaussian primitives, enabling each Gaussian to express multiple colors during rendering. 
In contrast, each vanilla Gaussian primitive in StyleSplat can only represent a single color, potentially leading to elliptical artifacts (e.g., unnatural elliptical distortions appearing at two ends of the wood dataset).
Additionally, by leveraging a combination of NNFM and CLIP losses, TexGS-VolVis's image-driven NPSE can capture local and global style patterns from the reference style image, producing stylization results that faithfully replicate the style patterns in the reference image.

\begin{table}[htb]
\caption{Average editing time (min), model size (MB), mean and max rendering time (ms) across image-driven NPSE datasets. 
We also report the average number of primitives used by StyleSplat and TexGS-VolVis.
The rendering resolution is 800$\times$800.}
\vspace{-0.1in}
\centering
\resizebox{3.0in}{!}{
\begin{tabular}{c|cccc}
   	     &editing & model & rendering time &  \#    \\ 
method   & time   & size  & (mean/max)   &   primitives\\ \hline
ARF & 5.9 &975 &23.75/ 25.11 &-- \\
StyleRF-VolVis & 8.4 &168 &243.61/348.58 &-- \\
StyleSplat & 6.9 &89 &6.11/10.27 &357,090 \\
TexGS-VolVis & 8.0 &223 &11.82/13.32 &129,857 \\ 
\end{tabular}
}
\label{tab:baselines-img-NPSE}
\end{table}

{\bf Quantitative evaluation.}
We report quantitative results in Table~\ref{tab:baselines-img-NPSE}. 
\hot{TexGS-VolVis is slightly slower than the ARF and StyleSplat because it uses both VGG and CLIP during optimization, while others use only VGG.}
Regarding model size, ARF requires significantly more storage due to its explicit feature grid. 
While TexGS-VolVis's model size is larger than StyleSplat's, it primarily stems from the need to store the texture attribute for each Gaussian, constituting over 90\% of the total model size.
Regarding rendering speed, TexGS-VolVis is significantly faster than StyleRF-VolVis and DVR (refer to Table~\ref{tab:dataset}). 
However, it is slightly slower than StyleSplat due to the additional lighting and texture mapping computations involved in the rendering process. 
After image-driven NPSE, TexGS-VolVis uses far fewer primitives compared to StyleSplat. 
Despite this, TexGS-VolVis produces more detail-rich stylization results, demonstrating that the texture attribute greatly enhances the expressive capacity of Gaussian primitives. 
This allows more complex style patterns to be represented with a limited number of primitives.
\begin{figure*}[htb]
 \begin{center}
 $\begin{array}{c@{\hspace{0.05in}}c@{\hspace{0.05in}}c@{\hspace{0.05in}}c@{\hspace{0.05in}}c}
 \includegraphics[width=0.175\linewidth]{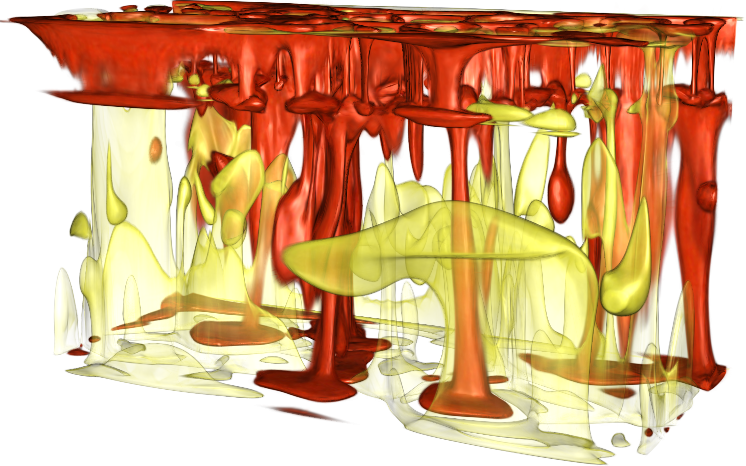}&
 \includegraphics[width=0.175\linewidth]{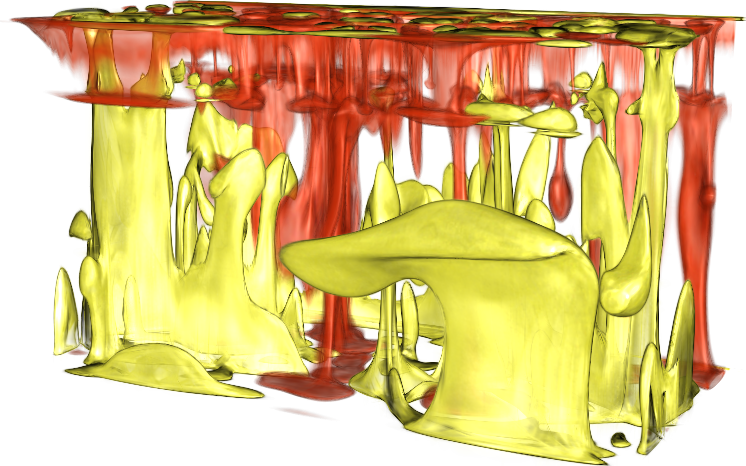}&
 \includegraphics[width=0.175\linewidth]{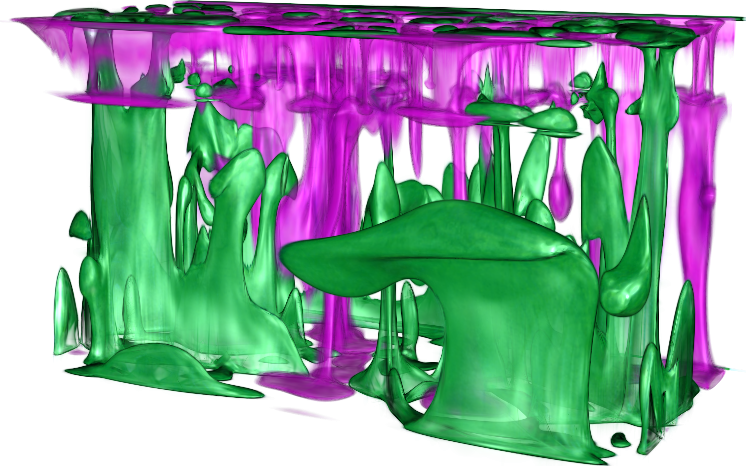}&
 \includegraphics[width=0.175\linewidth]{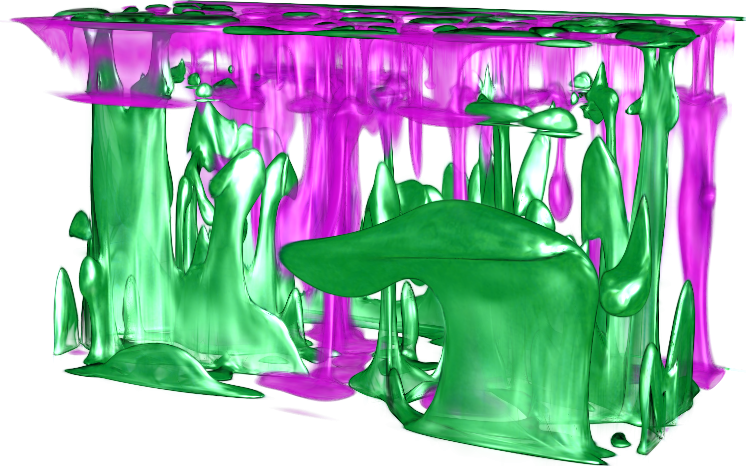}&
\includegraphics[width=0.175\linewidth]{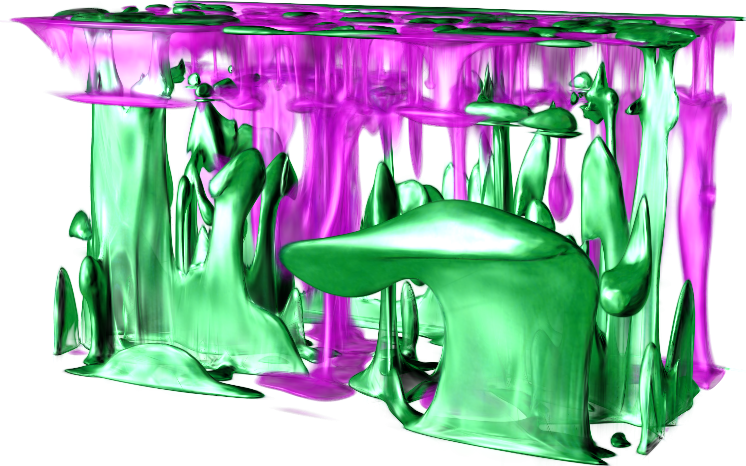}\\
\mbox{\footnotesize (a) selection of basic TFs} & \mbox{\footnotesize (b): (a) + opacity change}& \mbox{\footnotesize (c): (b) + color change} & \mbox{\footnotesize (d): (c) + light mag.\ change} & \mbox{\footnotesize (e): (d) + light dir.\ change}\\
 \includegraphics[width=0.175\linewidth]{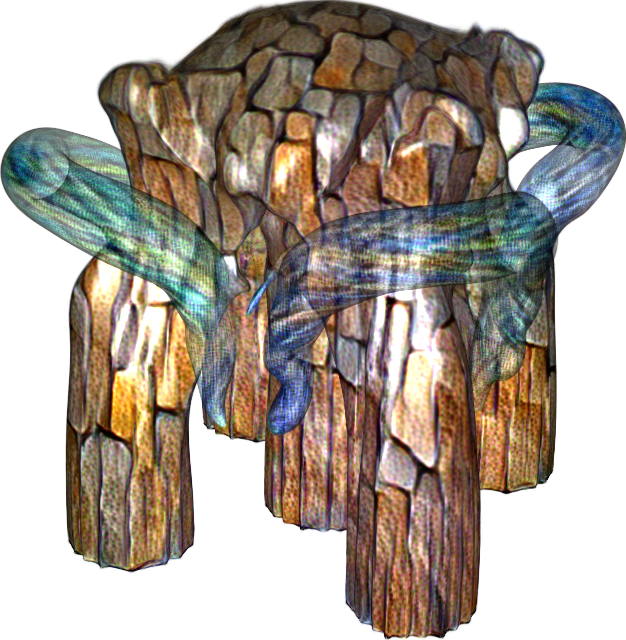}&
  \includegraphics[width=0.175\linewidth]{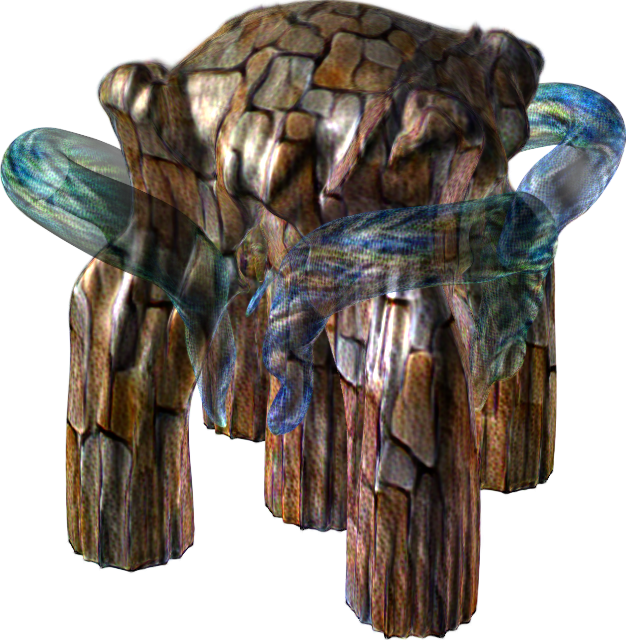}&
  \includegraphics[width=0.175\linewidth]{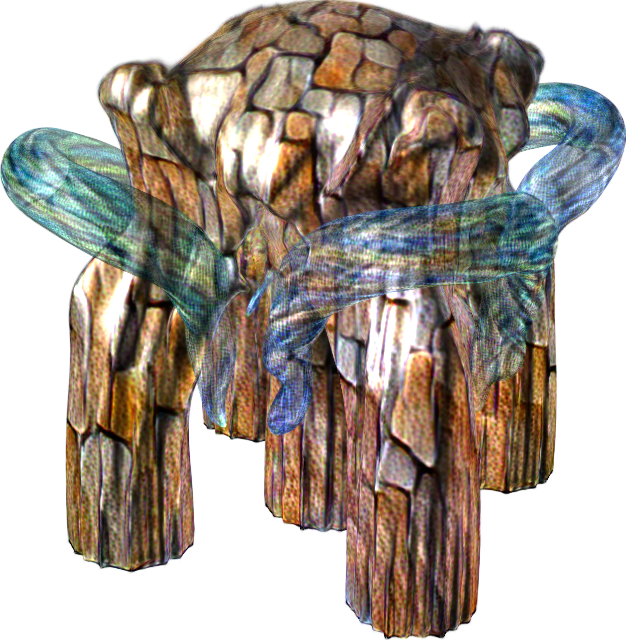}&
 \includegraphics[width=0.175\linewidth]{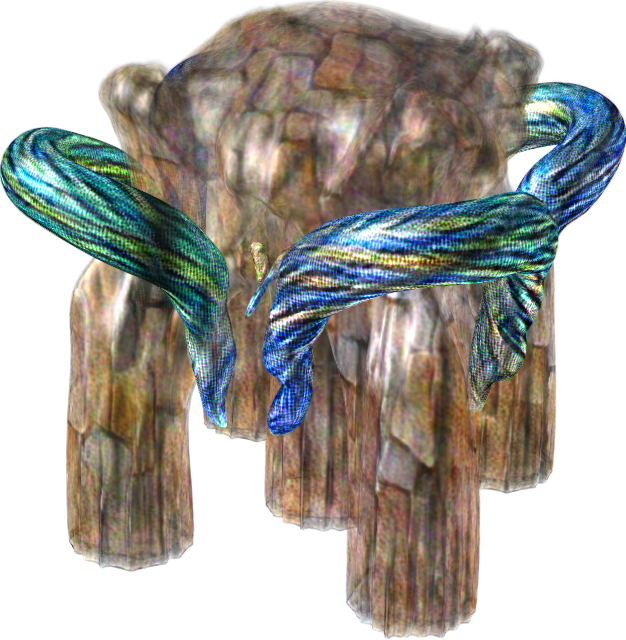}&
\includegraphics[width=0.175\linewidth]{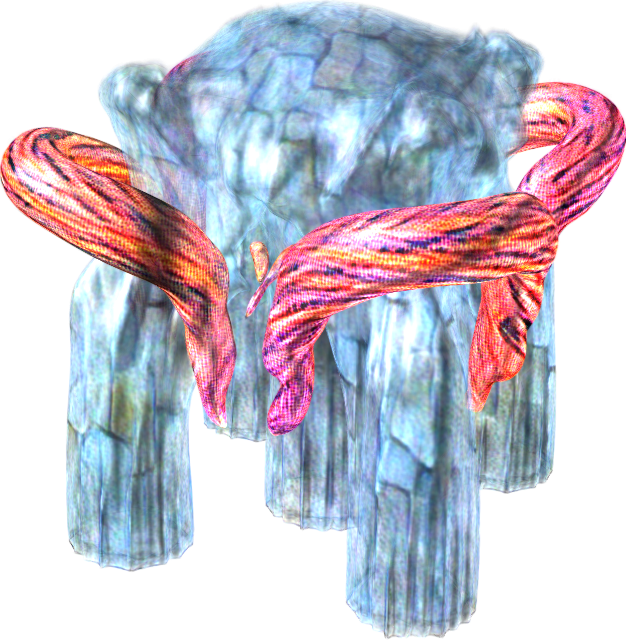}\\
\mbox{\footnotesize (f) selection of basic TFs} & \mbox{\footnotesize (g): (f) + light dir.\ change }& \mbox{\footnotesize (h): (g) + light mag.\ change} & \mbox{\footnotesize (i): (h) + opacity change} & \mbox{\footnotesize (j): (i) + color change}\\
\end{array}$
\end{center}
\vspace{-.25in} 
\caption{Iterative scene editing using the composed TexGS-VolVis model on the mantle and five jets datasets. 
In both cases, the light source shifts from the front to the left.} 
\label{fig:PSE}
\end{figure*}

\begin{figure}[htb]
 \begin{center}
 $\begin{array}{c@{\hspace{0.0in}}c}
 \includegraphics[width=0.485\linewidth]{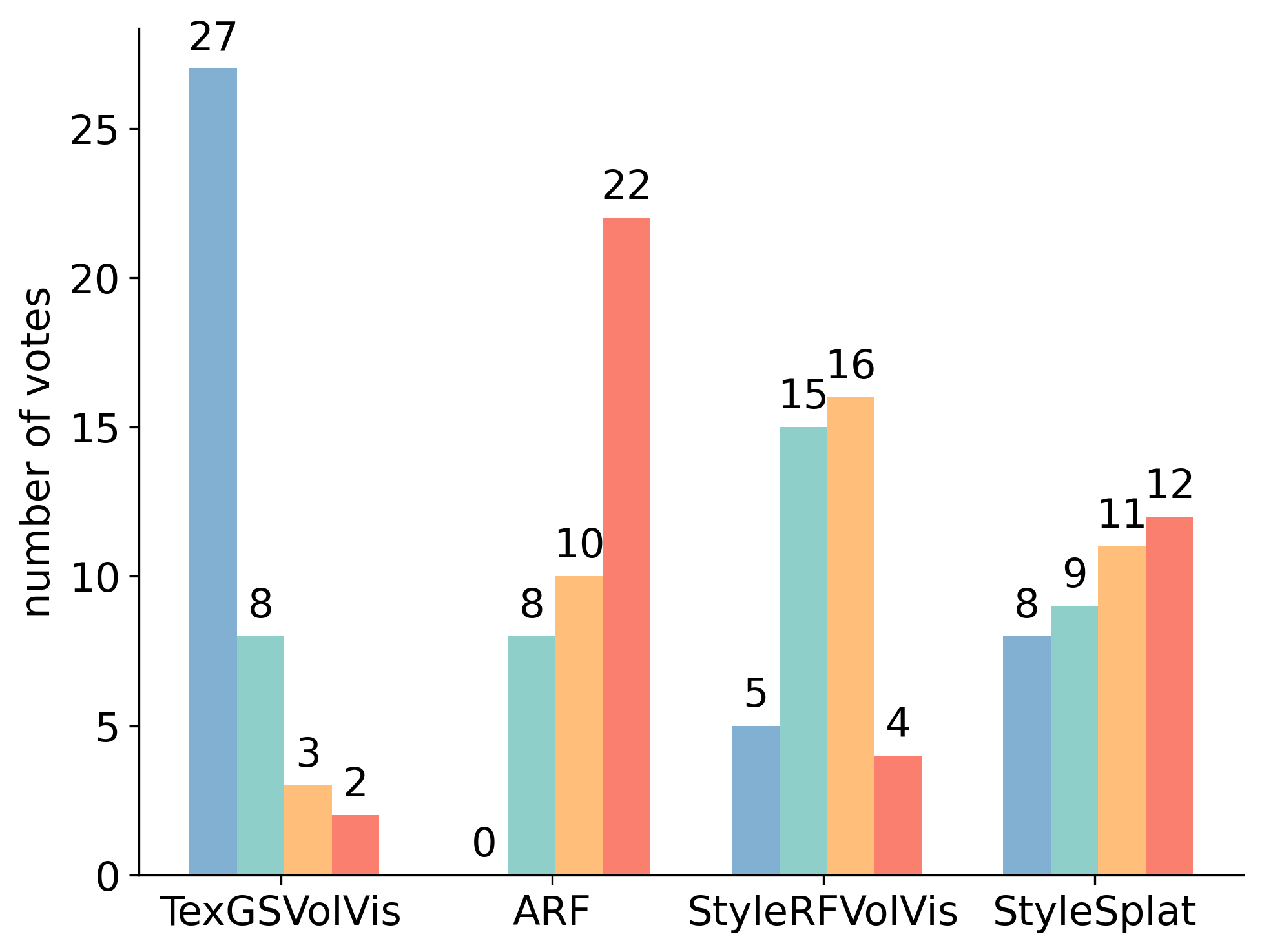}&
 \includegraphics[width=0.485\linewidth]{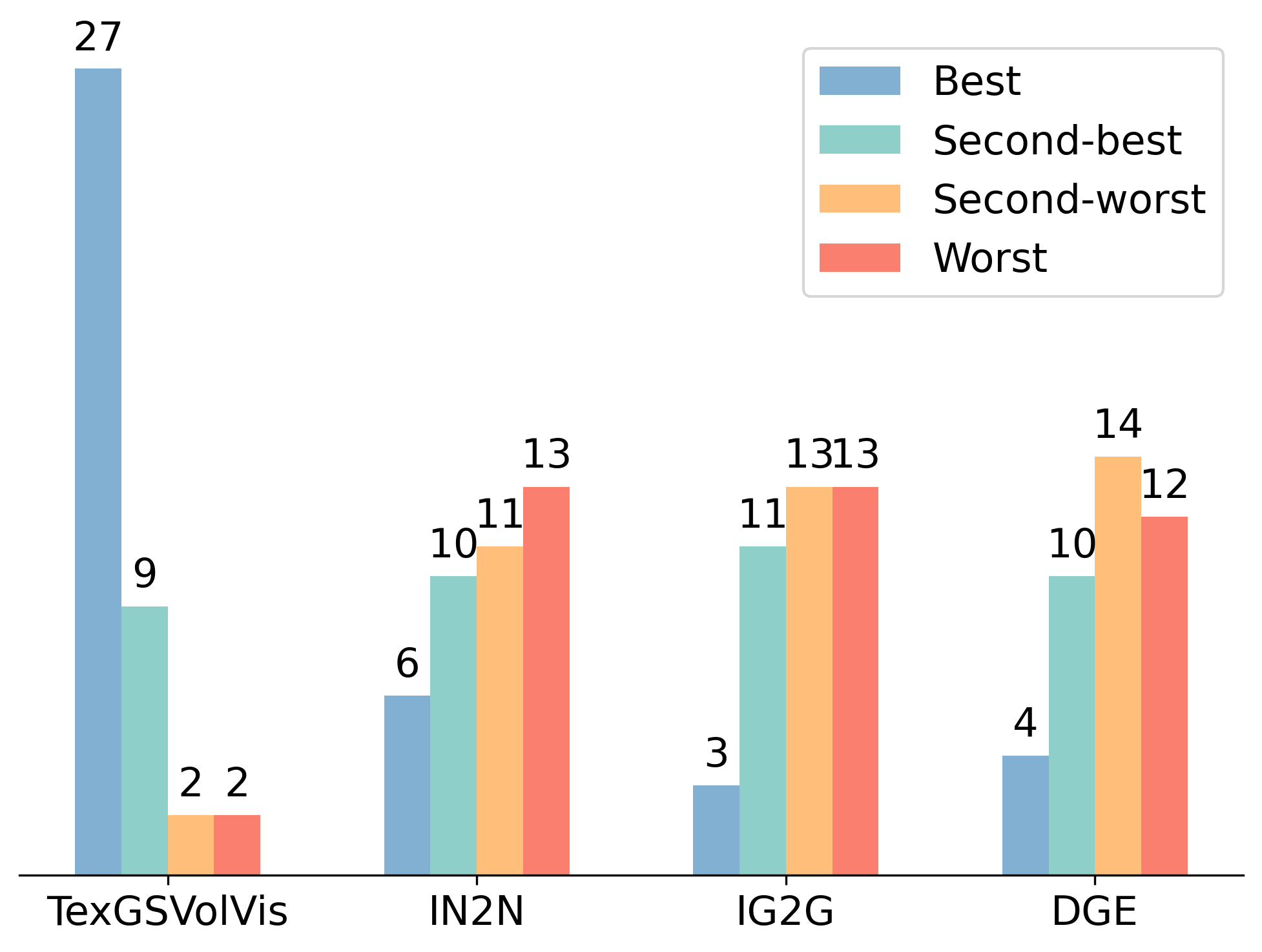}\\
\mbox{\footnotesize (a) image-driven NPSE} & \mbox{\footnotesize (b) text-driven NPSE} 
\end{array}$
\end{center}
\vspace{-.25in} 
\caption{The votes from ten participants ranking the stylization outcomes of image-driven NPSE (Figure~\ref{fig:baseline-img-NPSE}) and text-driven (Figure~\ref{fig:baseline-text-NPSE}) NPSE.} 
\label{fig:user-study}
\end{figure}

\vspace{-0.05in}
\subsection{Text-Driven NPSE}

{\bf Baselines and metrics.}
We compare TexGS-VolVis with three scene representation methods for text-guided NPSE using the IP2P~\cite{brooks-InstructPix2Pix} model:
\begin{myitemize}
\vspace{-0.05in}
\item IN2N~\cite{Haque-ICCV23} is a straightforward NeRF-based method that achieves stylization by iteratively updating the entire training dataset with 2D edits from IP2P (guided by text instructions) and refining the NeRF model through repeated training cycles to maintain 3D consistency.
\item Instruct-GS2GS (IG2G)~\cite{Vachha-igs2gs24} is a 3DGS-based method that follows the same pipeline as IN2N but utilizes 3DGS instead of a NeRF model to represent the scene for efficient rendering and optimization. 
\item DGE~\cite{Chen-DGE} is a 3DGS-based method designed for time-efficient and text-faithful stylization. 
It incorporates epipolar constraints into IP2P to perform multi-view consistent 2D edits on a small subset of training images. 
This is followed by directly optimizing the 3DGS model to align with the multi-view-consistent edited views.
\vspace{-0.05in}
\end{myitemize}
\noindent Note that the original IN2N uses an MLP network to represent the NeRF, which lacks composability and cannot combine independently stylized basic scenes. 
Therefore, to facilitate a fair comparison between IN2N and other methods, we replace the MLP representation with Plenoxels. 
We follow the common practice in previous work~\cite{Haque-ICCV23, Chen-DGE} by evaluating the alignment of text-driven NPSE scenes and target text prompts using the {\em CLIP similarity} (the cosine distance between the text and rendered image embeddings in CLIP space) and the {\em CLIP directional similarity} (the cosine distance between the image and text editing directions in CLIP space). 
These evaluations are performed under 181 novel views, uniformly sampled along a camera path with polar and azimuthal angles ranging from (-$90^{\circ}$, -$180^{\circ}$) to ($90^{\circ}$, $180^{\circ}$).

{\bf Qualitative evaluation.}
In Figure~\ref{fig:baseline-text-NPSE}, we compare the visual quality of the stylization results generated by baseline methods and TexGS-VolVis using the beetle, chameleon, engine, and ionization datasets. 
The rendered images show that TexGS-VolVis produces more detailed and accurate 3D edits that align closely with the input text prompts. 
For example, in the engine dataset, only TexGS-VolVis successfully stylizes the blue part with tiny dots resembling a stipple drawing, while other methods fail to replicate this effect. 
Additionally, the rendering from DGE exhibits noticeable elliptical artifacts. 
This occurs because IP2P's editing results are not always geometry-consistent with the scene content. 
DGE then updates its model on a small subset of multi-view images, which amplifies the impact of these inaccuracies.
Although TexGS-VolVis uses fewer multi-view images for text-driven NPSE than DGE, our approach keeps the geometry of the Gaussian primitives fixed while updating only the texture attribute for appearance editing. 
This allows for geometry-consistent editing without sacrificing expressiveness. 
In contrast, IN2N and IG2G rely on all training views to maintain geometry consistency and mitigate the effects of some inaccurate IP2P edits. 
However, due to these methods' lack of an expressive texture attribute, their editing results often exhibit low-frequency appearances and lack the detailed style patterns seen in TexGS-VolVis.

\begin{table}[htb]
\caption{Average CLIP similarity and CLIP directional similarity, editing time (min), mean and max rendering time (ms), as well as model size (MB) for text-driven NPSE datasets. 
We also report the average number of primitives used by IG2G, DGE, and TexGS-VolVis.
The rendering resolution is 512$\times$512.}
\vspace{-0.1in}
\centering
\resizebox{\columnwidth}{!}{
\begin{tabular}{c|cc|cc|cc}
   	     & CLIP     &CLIP directional  & editing   & rendering time & model & \#   \\ 
method & similarity$\uparrow$ &similarity$\uparrow$  	&  time   & (mean/max) & size & primitives \\ \hline
IN2N &0.343&0.043&44.69 &9.33/10.72 &1,150&--  \\
IG2G &0.363&0.134&38.3&3.61/5.16 &57&231,189 \\
DGE &0.364&0.133 &7.2&4.14/5.06 &87&351,917\\
TexGS-VolVis & 0.366 & 0.174 &1.6&7.21/9.93 &221&108,119 \\ 
\end{tabular}
}
\label{tab:baselines-text-NPSE}
\end{table}

{\bf Quantitative evaluation.}
We report quantitative results in Table~\ref{tab:baselines-text-NPSE}. 
TexGS-VolVis achieves the highest average CLIP scores across all datasets. 
Similar to the image-driven NPSE results, TexGS-VolVis uses far fewer primitives than other GS-based methods like IG2G and DGE, but it does experience slightly slower rendering speeds and incur a larger model size. 
However, the editing time for TexGS-VolVis is notably faster than other methods. 
This can be attributed to the fact that, compared to optimizing Plenoxels or Gaussian primitives, the inference process of IP2P typically requires more time. 
Unlike the baseline methods that employ an iterative updating strategy—repeatedly running IP2P over many multi-view images—TexGS-VolVis uses a one-time update strategy. 
It applies IP2P editing only once on a small subset of multi-view images, resulting in a much shorter overall editing time.
\hot{On the other hand, within TexGS-VolVis, image-driven NPSE requires frequently applying VGG and CLIP models during optimization, resulting in slower editing speeds compared to text-driven NPSE.}

\begin{figure*}[htb]
 \begin{center}
 $\begin{array}{c@{\hspace{0.025in}}c@{\hspace{0.025in}}c@{\hspace{0.025in}}c}
 \includegraphics[height=1.15in]{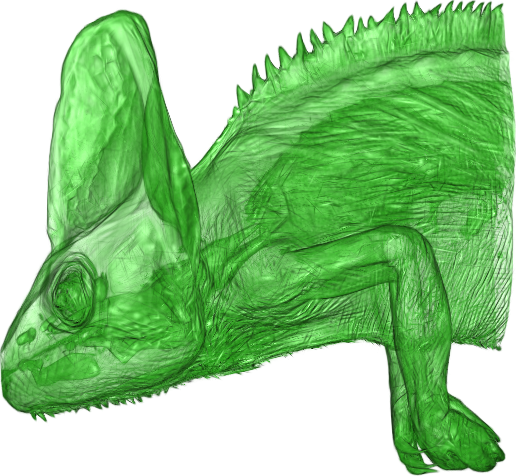}&
  \includegraphics[height=1.15in]{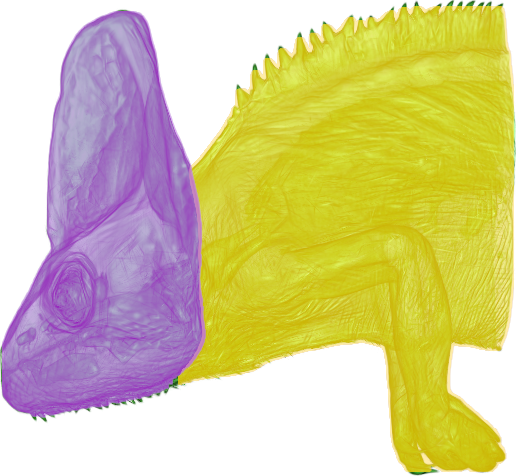}&
  \includegraphics[height=1.15in]{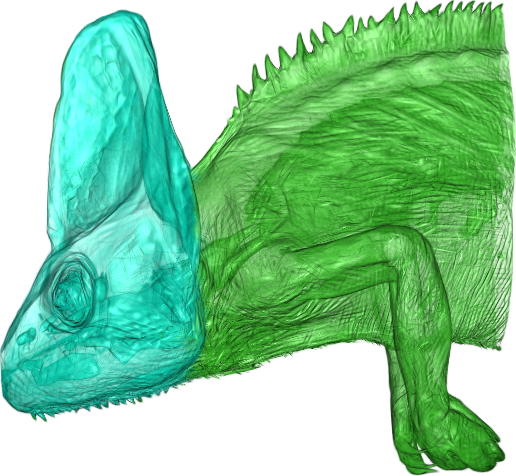}&
 \includegraphics[height=1.15in]{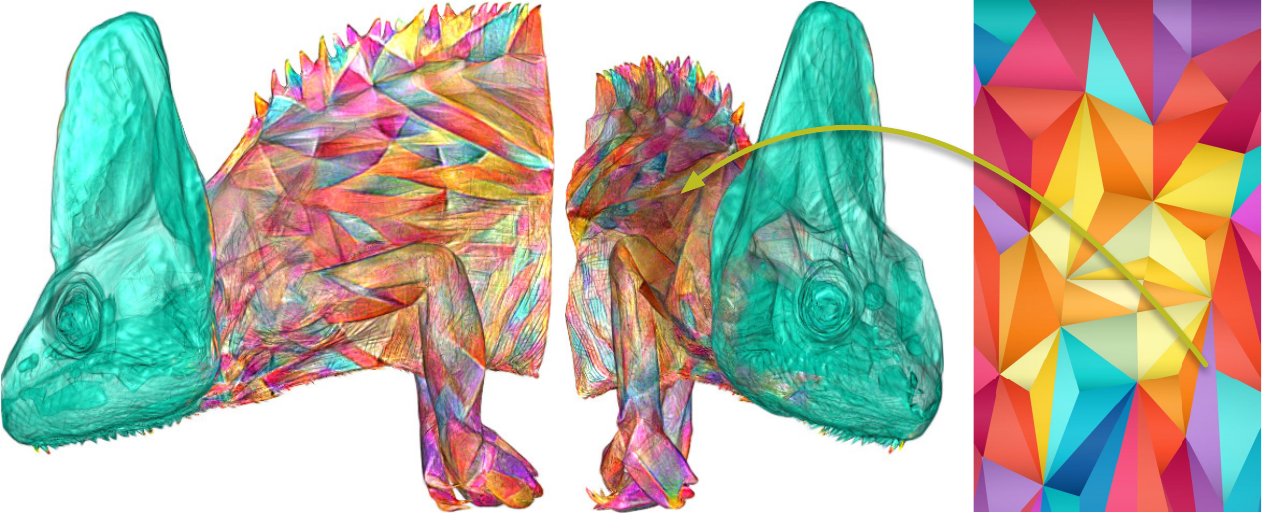}\\
  \includegraphics[height=1.15in]{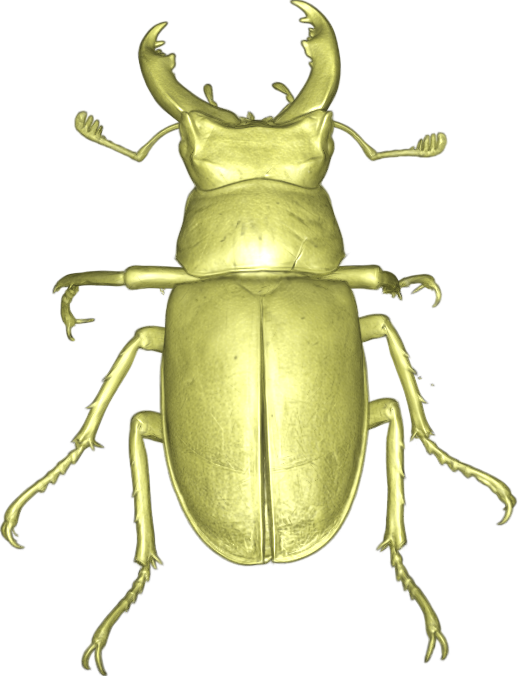}&
 \includegraphics[height=1.15in]{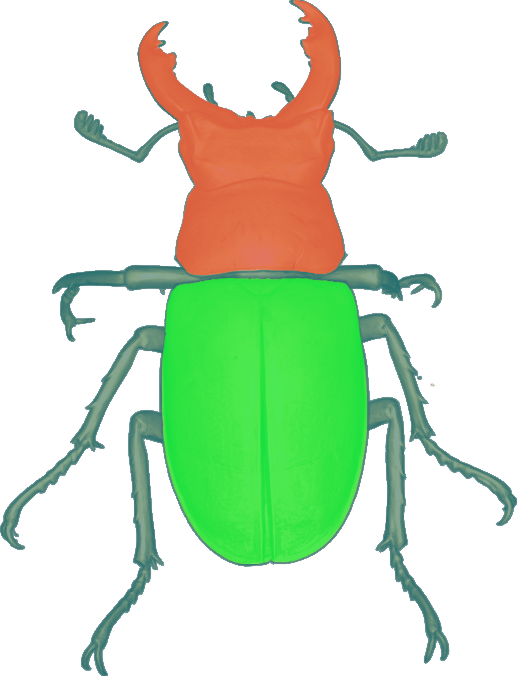}&
 \includegraphics[height=1.15in]{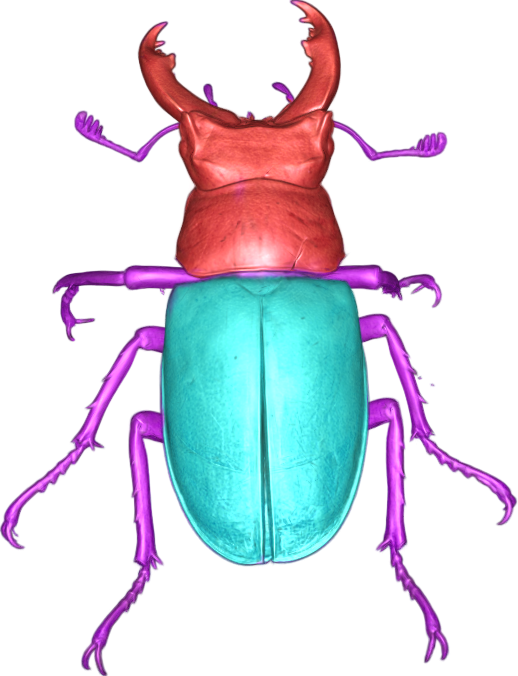}&
\includegraphics[height=1.15in]{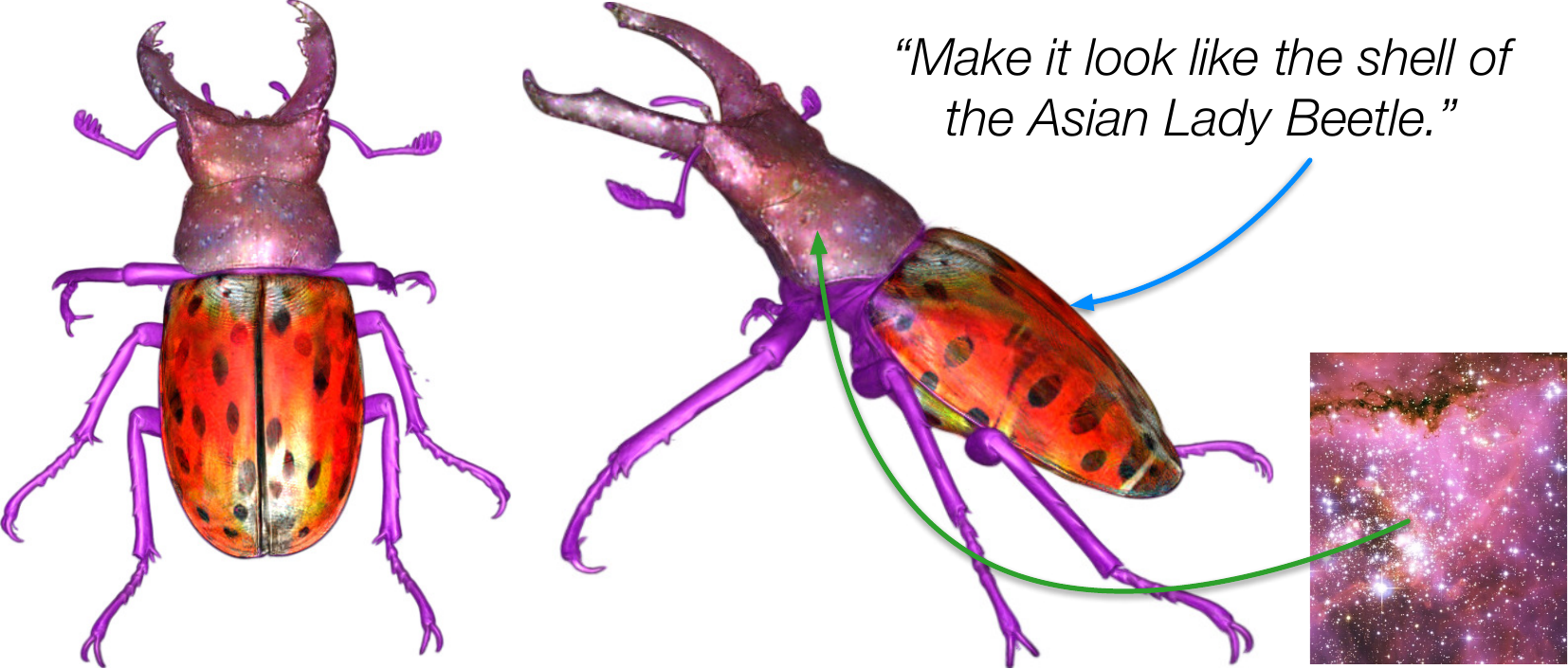}\\
\mbox{\footnotesize (a) original scene} & \mbox{\footnotesize (b) 2D segmentation mask}& \mbox{\footnotesize (c) PSE result} & \mbox{\footnotesize (d) PSE+NPSE result}\\
\end{array}$
\end{center}
\vspace{-.25in} 
\caption{Partial scene editing using 2D-lift-3D segmentation on one basic scene from the chameleon and beetle datasets. 
Given (a) a basic scene, utilizing (b) the 2D segmentation mask generated by the SAM model on a reference view, we can lift the 2D mask to a 3D mask to segment the scene into distinct parts. 
Various edits can be applied to each segment, as shown in (c) and (d).} 
\label{fig:2DLift3D}
\end{figure*}

\vspace{-0.05in}
\subsection{User Study}

To further evaluate the perceived quality of NPSE, we conducted a user study to measure participant preferences for different NPSE results, as shown in Figures~\ref{fig:baseline-img-NPSE}~and~\ref{fig:baseline-text-NPSE}, following the University's IRB protocol. 
We recruited ten students ranging from undergraduates to Master's and PhD candidates. 
Before the study, participants were given a brief overview of the procedure. 
Each result is presented in full-screen mode, displaying the original VolVis scene and the reference images or text prompts at the top. 
The stylization results of TexGS-VolVis and the baseline methods are shown at the bottom. 
All scenes are presented as video clips, with the 3D scene gradually changing as the camera view moves. 
For each case, the stylization video results of TexGS-VolVis and the baseline methods were randomly arranged and labeled. 
Participants were asked to rank the stylization outcomes from best to worst without ties. 
They were instructed to consider several factors, such as overall impression, content preservation, style application, and visual aspects like color, opacity, and lighting. 
Participants could determine the relative importance of each factor but were asked to apply their chosen criteria consistently throughout the study.
The results, given in Figure~\ref{fig:user-study}, show that TexGS-VolVis outperforms other baseline approaches in both image- and text-driven NPSE.
Please refer to \hot{Section 4 of the appendix} for a detailed breakdown of voting results for all methods across each dataset.

\vspace{-0.05in}
\subsection{PSE and 2D-Lift-3D Segmentation}

TexGS-VolVis supports real-time PSE during inference, allowing for color, opacity, and lighting adjustments to the represented scene. 
Like NPSE, different PSE effects can be applied to various basic scenes defined by their respective basic TFs. 
Figure~\ref{fig:PSE} illustrates the iterative PSE results for the mantle and five jets datasets. 
In contrast to StyleRF-VolVis~\cite{Tang-VIS24}, which only supports color editing before stylization and cannot adjust light direction after training, TexGS-VolVis uniformly represents both non-stylized and stylized VolVis scenes using \hot{the textures attribute}. 
This enables color editing even after stylization. 
Furthermore, by optimizing shading attributes and employing the Blinn-Phong shading model, TexGS-VolVis supports lighting magnitude and direction adjustment.

In StyleRF-VolVis, users cannot apply distinct edits to different parts of a basic scene. 
To enhance editing granularity and flexibility, we introduce 2D-lift-3D segmentation for partial scene editing. 
Figure~\ref{fig:2DLift3D} demonstrates the partial scene editing results on the chameleon and beetle datasets. 
We perform 3D segmentation by segmenting Gaussian primitives within the basic scene based on the 2D segmentation results generated by the SAM model and user-specified point prompts. 
After obtaining the 3D segmentations, we can decompose a basic scene into different parts and independently apply various editing effects to each part. 
This significantly enhances editing control and flexibility.

\begin{figure}[htb]
 \begin{center}
 $\begin{array}{c@{\hspace{0.025in}}c}
  \includegraphics[width=0.475\linewidth]{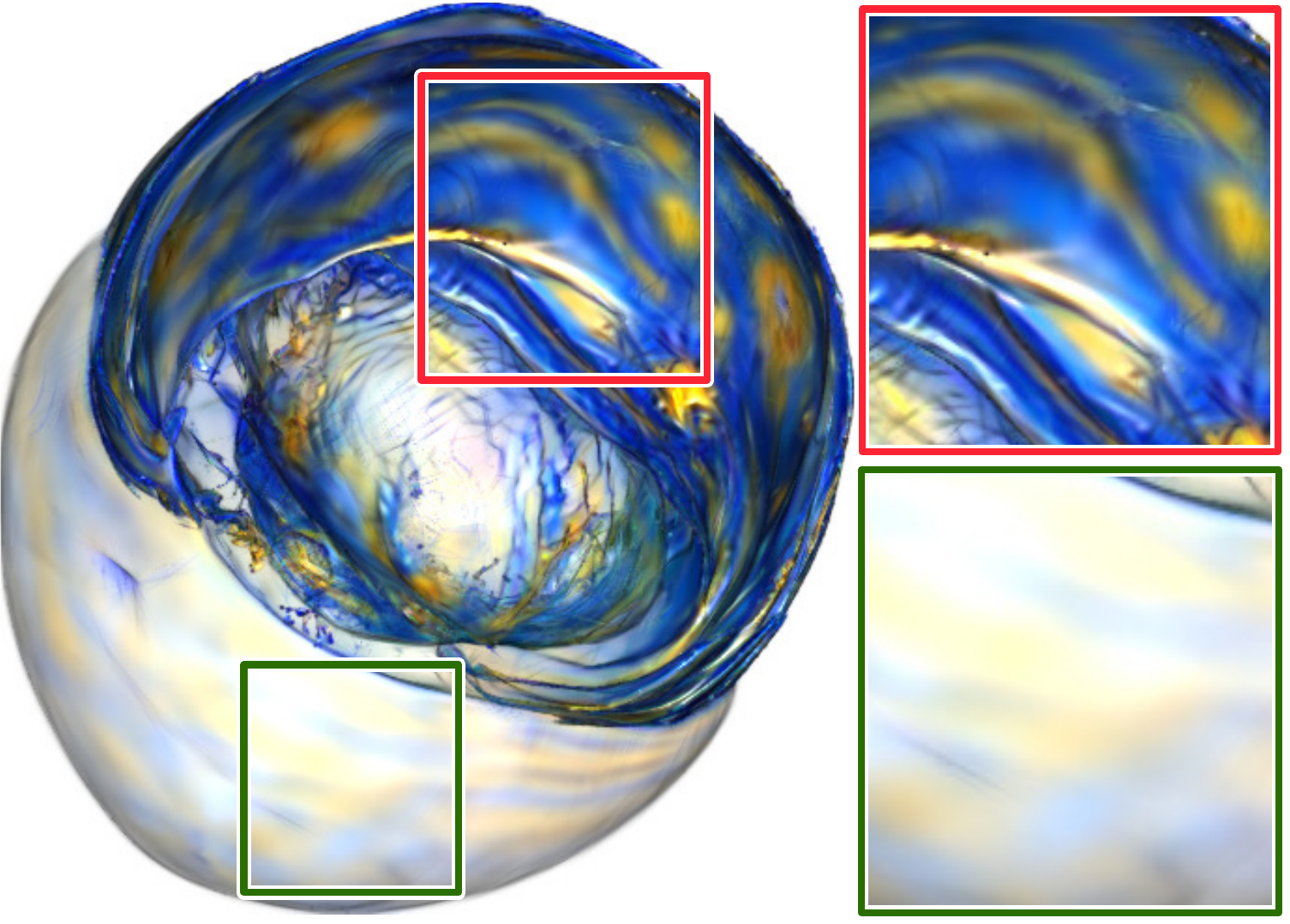}&
 \includegraphics[width=0.475\linewidth]{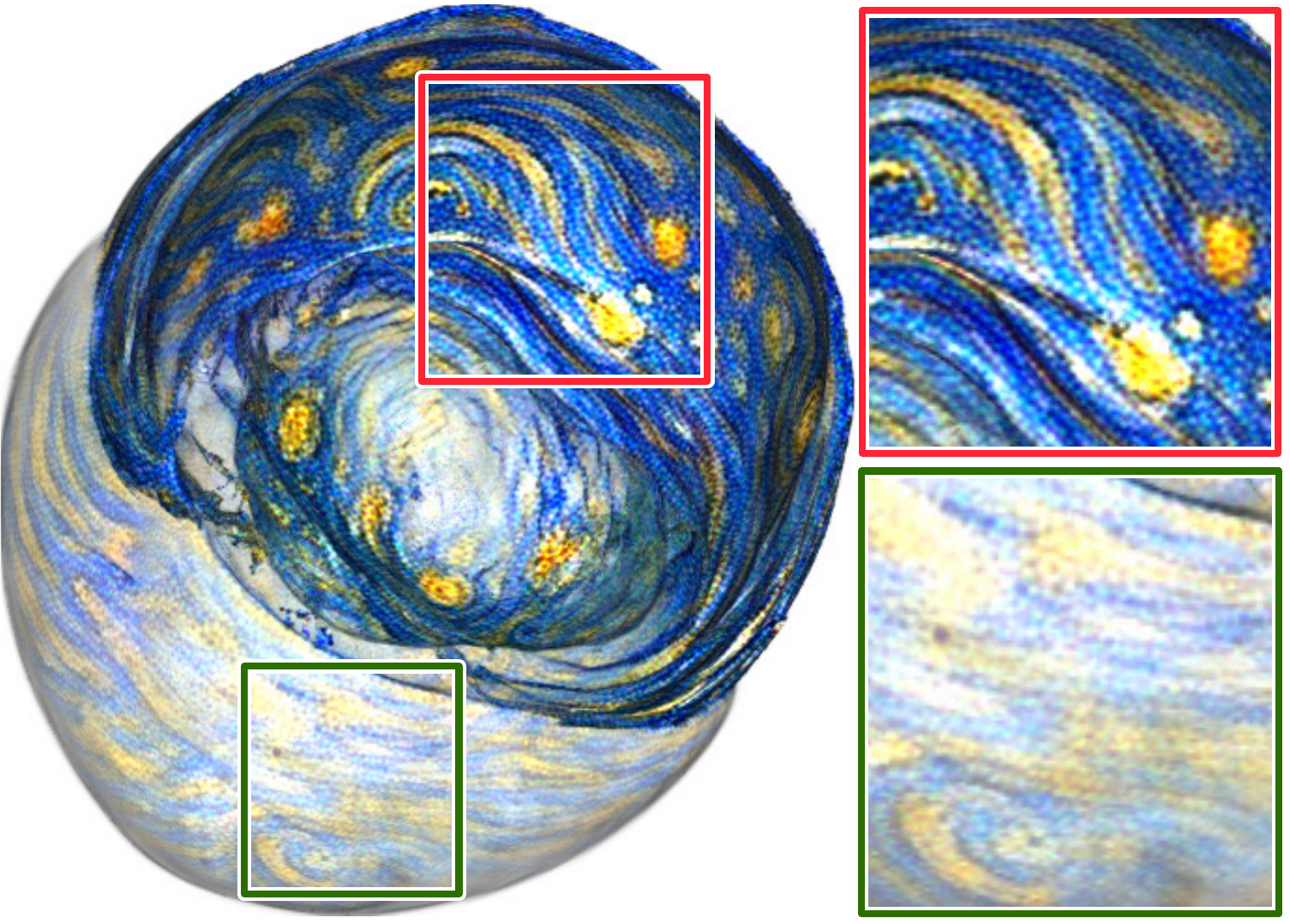}\\
 \includegraphics[width=0.475\linewidth]{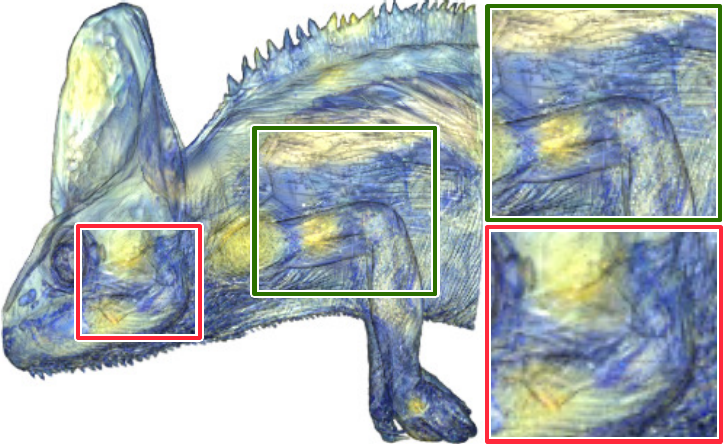}&
 \includegraphics[width=0.475\linewidth]{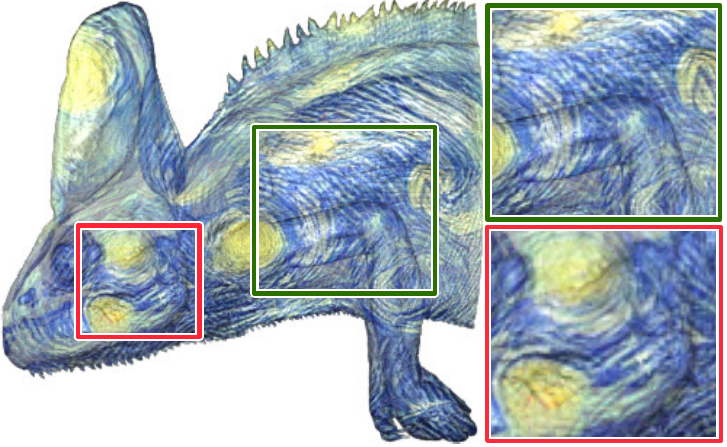}\\
\mbox{\footnotesize (a) w/o texture} & \mbox{\footnotesize (b) w/ texture} 
\end{array}$
\end{center}
\vspace{-.25in} 
\caption{\hot{Comparison of TexGS-VolVis's image-driven (top) and text-driven (bottom) NPSE results with and without the texture attribute for the supernova in Figure~\ref{fig:baseline-img-NPSE} and the chameleon's green part in Figure~\ref{fig:baseline-text-NPSE}, respectively. The ones with the texture attribute produce more expressive style patterns.}} 
\label{fig:ablation-texture}
\end{figure}

\vspace{-0.15in}
\subsection{Texture Attribute in NPSE}
\label{subsec:ablation}

To emphasize the importance of the texture attribute in achieving high-quality NPSE, we perform \hot{image and text-driven NPSE on the yellow part of the supernova and the green part of the chameleon—with and without the texture attribute. 
The same image and text prompts used in Figures~\ref{fig:baseline-img-NPSE}~and~\ref{fig:baseline-text-NPSE} are applied in this comparison.}
For the version of TexGS-VolVis without the texture attribute, we edit the appearance by updating the view-independent color $\mathbf{c}_{\ind}$ within each primitive. 
As shown in Figure~\ref{fig:ablation-texture}, TexGS-VolVis with the texture attribute produces significantly richer style patterns, highlighting its crucial role in enhancing the expressive capabilities of Gaussian primitives.

\vspace{-0.05in}
\subsection{Limitations}

Despite the effectiveness of TexGS-VolVis in achieving flexible, high-quality NPSE with real-time rendering, it still has several limitations.
First, the texture attribute in TexGS-VolVis can occupy significant storage space \hot{(refer to Section 4 in the appendix)}, especially as the number of texels increases, resulting in a larger model size than other GS methods.
Second, the text-based editing capability of TexGS-VolVis heavily relies on the expressiveness of the IP2P model, which is based on a pretrained stable diffusion model. 
Inaccurate or vague text prompts can lead to erroneous editing results from IP2P, negatively affecting the NPSE outcomes of TexGS-VolVis.
Third, the 2D-lift-3D segmentation method we use cannot immediately generate a 3D segmentation from the 2D mask created by user-specified point prompts on a reference view. 
This can be inefficient for certain VolVis scenes that require multiple rounds of segmentation.

\vspace{-0.1in}
\section{Conclusions and Future Work}

We have introduced TexGS-VolVis, the first 2DGS-based framework in VolVis aimed at expressive scene representation and editing. 
Unlike the previous work, StyleRF-VolVis, TexGS-VolVis achieves higher-quality image-driven NPSE results, provides a more intuitive text-driven NPSE, and enables more flexible partial editing via 2D-lift-3D segmentation. 
TexGS-VolVis utilizes 2D Gaussian primitives enhanced with shading and texture attributes, allowing for real-time rendering, scene relighting with the Blinn-Phong shading model, and expressive appearance representation. 
We compared our method with other image-driven (ARF, StyleRF-VolVis, and StyleSplat) and text-driven (IN2N, IG2G, and DGE) NPSE solutions through qualitative and quantitative evaluations, demonstrating its superior performance.

In future work, we aim to improve TexGS-VolVis in three directions. 
First, we plan to compress the model of TexGS-VolVis by applying advanced compression techniques~\cite{Niedermayr-CVPR24, Fan-NIPS24} to make Gaussian attributes more compact. 
\hot{
Second, we plan to investigate instant style transfer methods~\cite{Liu-Siggraph24} to eliminate the need for per-scene optimization, thereby improving the system's responsiveness for interactive applications.
Third, we aim to extend TexGS-VolVis to support time-varying volumetric scene representation and editing, which will require the development of time-dependent Gaussian primitives~\cite{Yao-VIS25} to address the unique challenges of modeling dynamic VolVis scenes.} 
Finally, TexGS-VolVis already demonstrates faster rendering speeds and more compact storage than conventional DVR on large-scale volume datasets, making it ideal for deployment on low-end local devices for real-time rendering. 
We envision integrating this approach into VR devices, enabling natural language interaction~\cite{Ai-VIS25} and offering users an immersive exploration and editing experience on large-scale volumetric datasets.

\vspace{-0.05in}
\acknowledgments{This research was supported in part by the U.S.\ National Science Foundation through grants IIS-1955395, IIS-2101696, OAC-2104158, and IIS-2401144, and the U.S.\ Department of Energy through grant DE-SC0023145. The authors would like to thank the anonymous reviewers for their insightful comments.}

\vspace{-0.05in}
\section*{Appendix}

\setcounter{section}{0}
\setcounter{figure}{0}
\setcounter{table}{0}

\section{Interactive Interface}

\begin{figure}[htb]
 \begin{center}
 \includegraphics[width=1.0\linewidth]{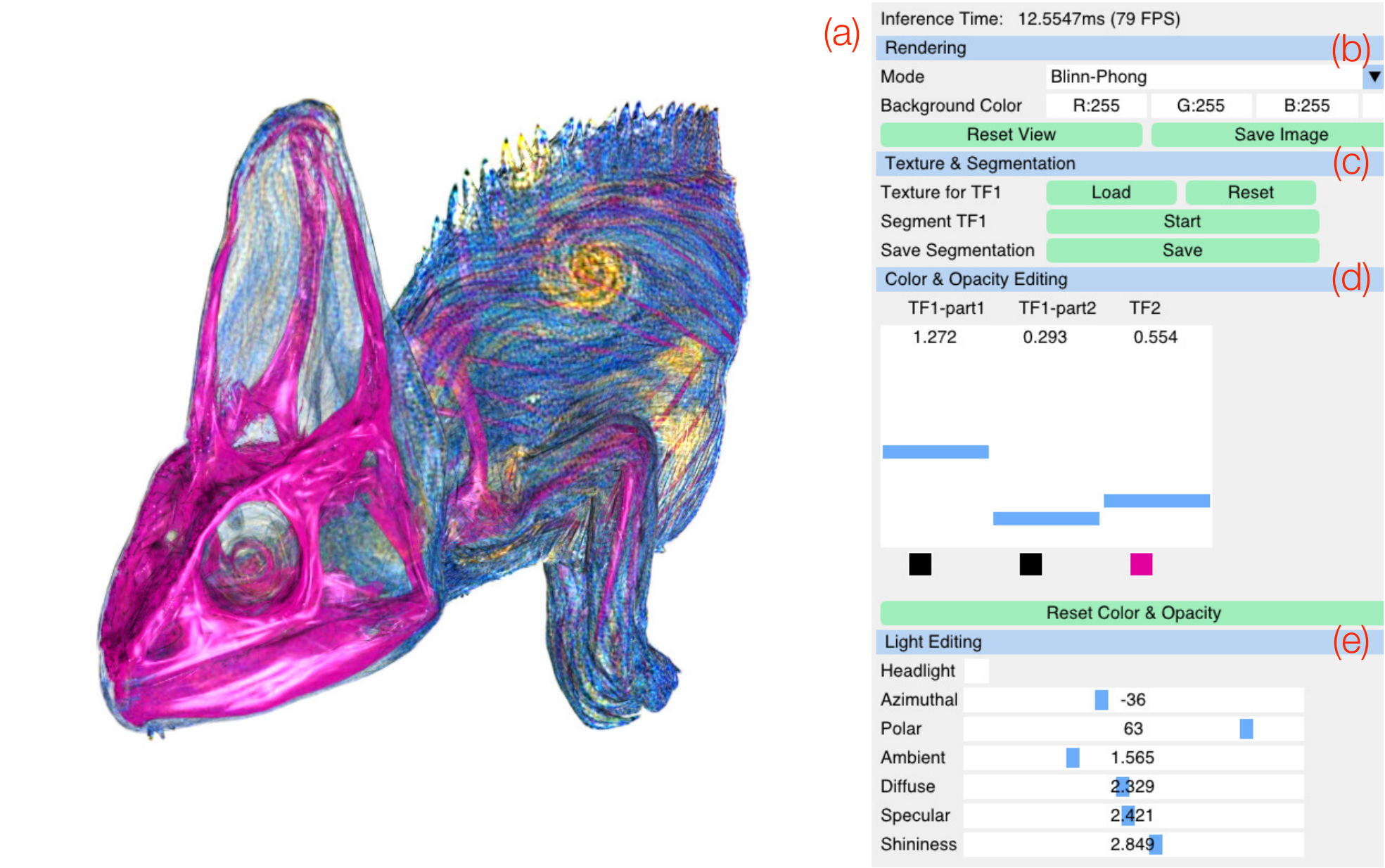}
\end{center}
\vspace{-.25in} 
\caption{A screenshot of TexGS-VolVis displaying the rendering results of the chameleon dataset.
(a) Interactive viewing area showcasing the rendering output.
(b) Options for selecting different rendering modes (e.g., Blinn-Phong shading, normal, and depth).
(c) Controls for loading edited textures, segmentation, and saving the model.
(d) Controls for adjusting color and opacity.
(e) Options for modifying lighting effects.} 
\label{fig:GUI}
\end{figure}

Figure~\ref{fig:GUI} presents our graphical user interface, which allows users to visualize and edit the VolVis scene. 
Thanks to the efficient rendering mechanism of 2DGS, our system supports real-time rendering (over 30 FPS), ensuring smooth and responsive user interaction within the interface. 
For a demonstration of the interactions, please refer to the accompanying video.

\vspace{-0.05in}
\section{Parameter Study}

\hot{
{\bf Evaluation of training views.}
To investigate the relationship between scene representation performance and the number of training views, we optimize TexGS-VolVis on the ionization dataset using different numbers of training images uniformly sampled via icosphere sampling.
As shown in Table~\ref{tab:ablation-training-view}, varying the number of training images does not significantly affect the fitting time or the number of primitives in the model.
However, reconstruction accuracy improves as more views are provided, while insufficient views lead to overfitting.
In our experiments, we use 162 training views, as increasing this number further yields only marginal gains in reconstruction quality.
}

\begin{table}[htb]
\caption{\hot{Average training and testing PSNR (dB), SSIM, fitting time (min), and number of primitives used for each basic scene of the ionization dataset under different numbers of training views.}}
\vspace{-0.1in}
\centering
\resizebox{\columnwidth}{!}{
\hot{
\begin{tabular}{c|cc|cc|cc}
\#  &\multicolumn{2}{c|}{training views} &\multicolumn{2}{c|}{testing views}&fitting&\#\\
views   & PSNR$\uparrow$   & SSIM$\uparrow$  & PSNR$\uparrow$   &   SSIM$\uparrow$ &time&primitives\\ \hline
42 &33.68&0.9849&29.33&0.9519 &5.6 &60,172 \\
92 &32.67&0.9764&30.95&0.9659 &5.5 &59,885 \\
162 &32.89& 0.9748 &31.64 & 0.9712 &5.6 &59,698 \\
252 &31.92 & 0.9721 &31.71& 0.9725 &5.7 &58,977 \\ 
\end{tabular}
}
}
\label{tab:ablation-training-view}
\end{table}

\begin{figure}[htb]
 \begin{center}
 $\begin{array}{c@{\hspace{0.025in}}c@{\hspace{0.025in}}c}
 \includegraphics[width=0.315\linewidth]{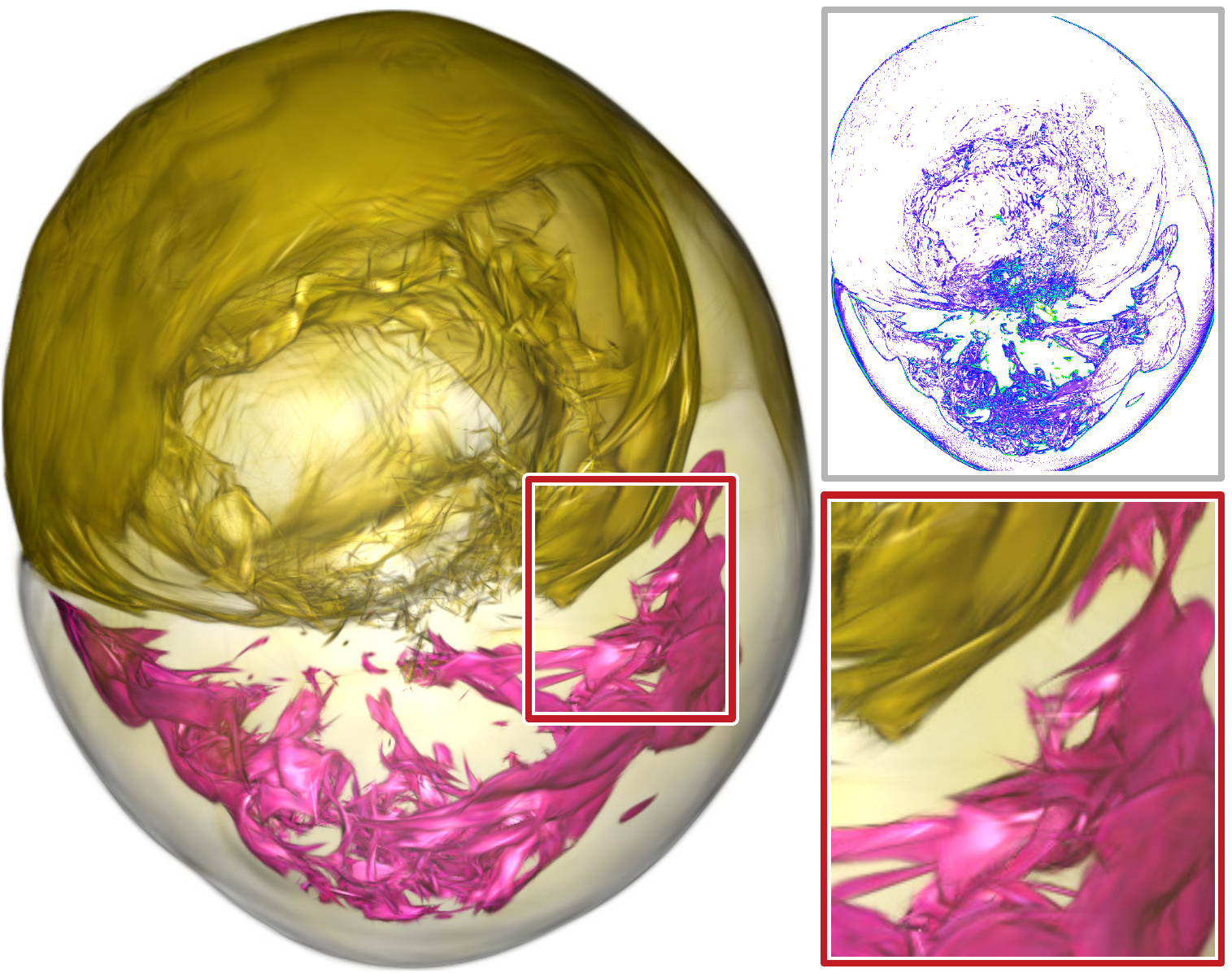}&
 \includegraphics[width=0.315\linewidth]{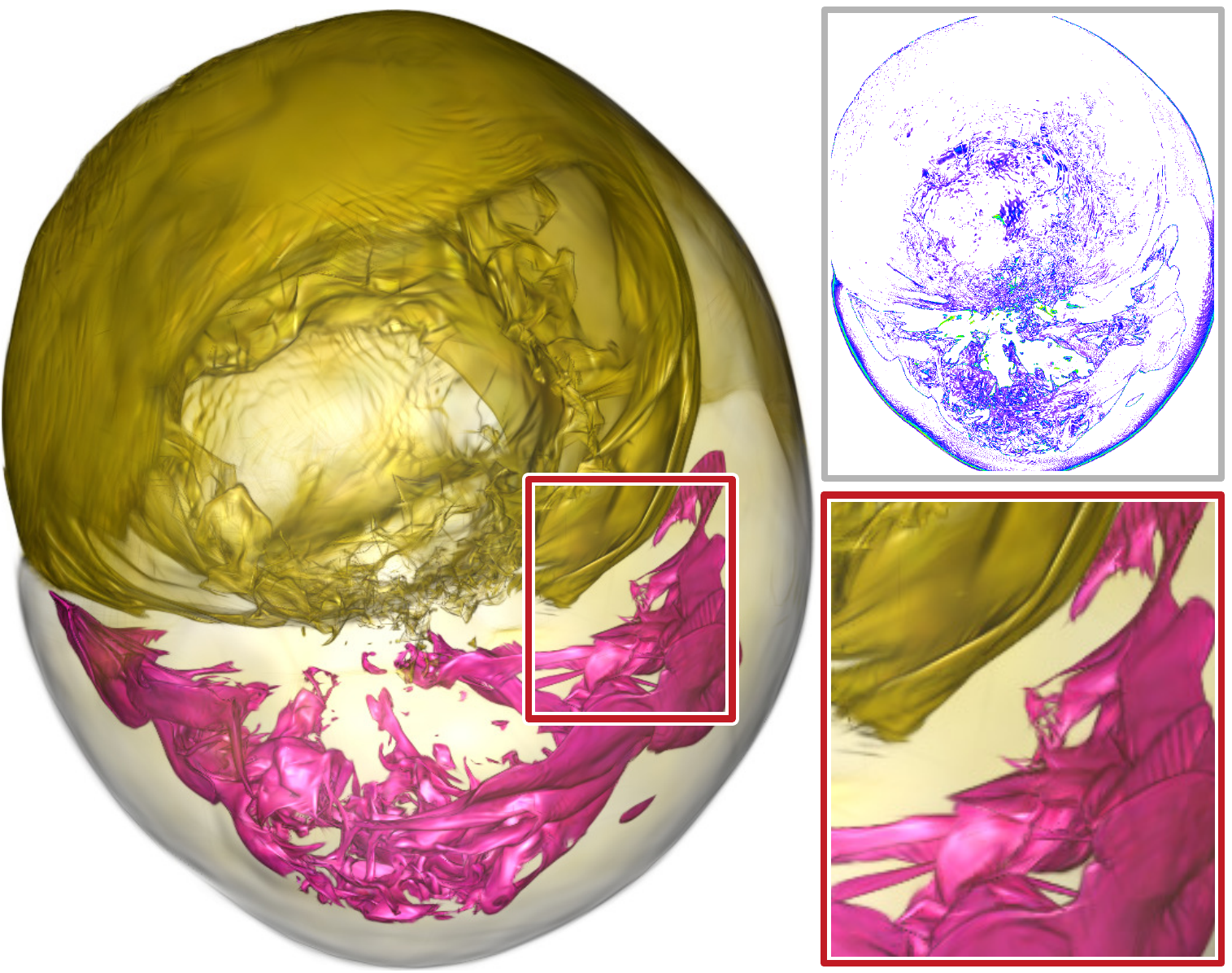}&
 \includegraphics[width=0.315\linewidth]{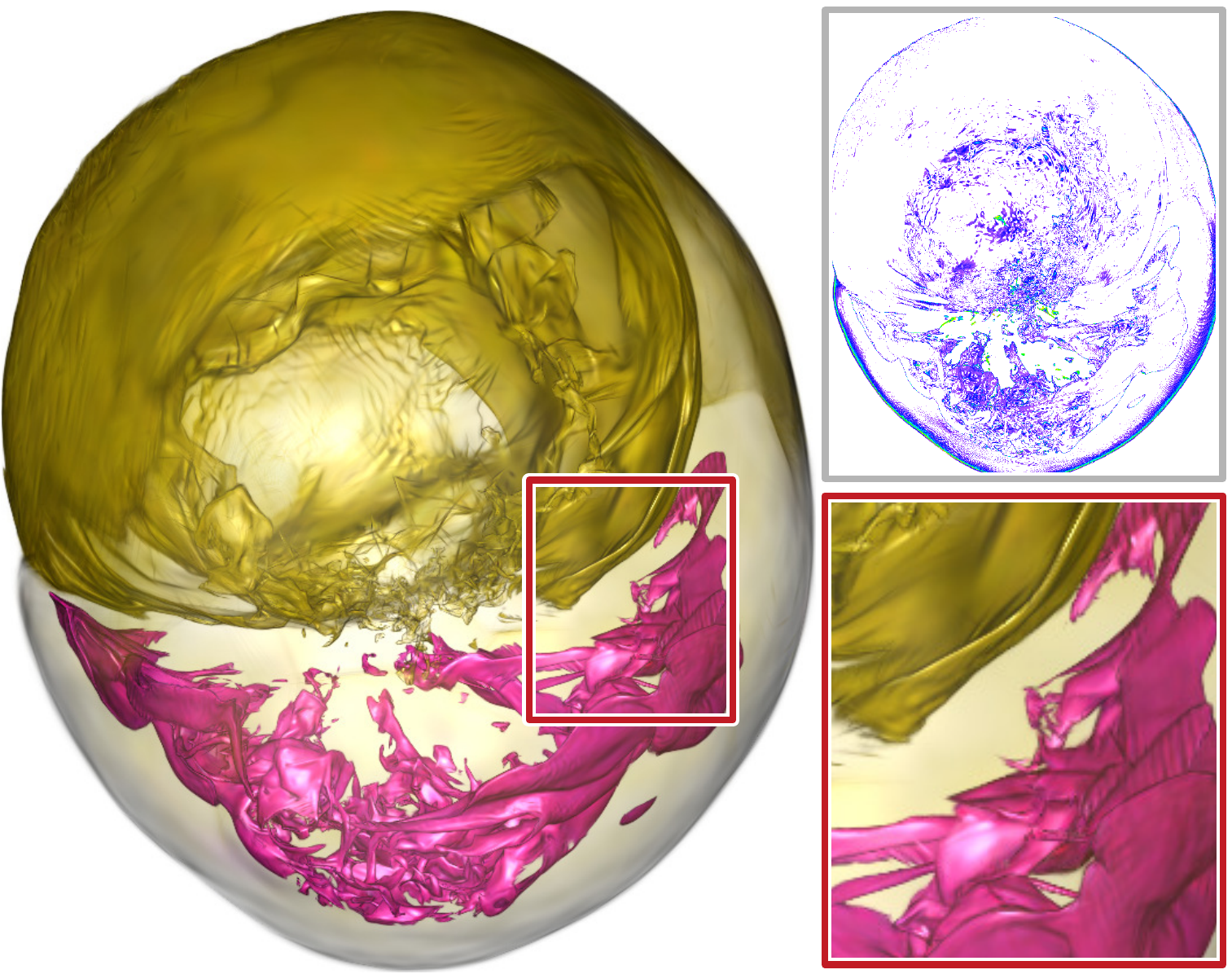}\\
\mbox{\footnotesize (a) 400$\times$400$\uparrow$} & \mbox{\footnotesize (b) 800$\times$800$\uparrow$} & \mbox{\footnotesize (c) 1200$\times$1200} 
\end{array}$
\end{center}
\vspace{-.25in} 
\caption{\hot{Rendering results of TexGS-VolVis on the supernova dataset at 1200$\times$1200 resolution, with models trained under different image resolutions. $\uparrow$ indicates that the model is trained on lower resolution images and rendered using higher resolution.}} 
\label{fig:ablation-img-res}
\end{figure}

\hot{
{\bf Evaluation of training image resolution.}
To examine how the training image resolution affects the TexGS-VolVis scene representation, we train our model on the supernova dataset with varying training image resolutions.
Table~\ref{tab:ablation-img-res} reports the reconstruction accuracy, fitting time, and number of primitives used.
Note that TexGS-VolVis rasterizes optimized primitives in 3D space, so the rendering resolution can be set independently of the training image resolution.

Figure~\ref{fig:ablation-img-res} shows rendering results with 1200$\times$1200 resolution using TexGS-VolVis models optimized under different training image resolutions. 
As shown in the zoomed-in regions and difference images, models trained at 400$\times$400 resolution produce elliptical artifacts and inaccurate reconstructions when rendered at higher resolutions, indicating that the primitive resolution is too low to capture fine-grained structural details. 
Models optimized at 800$\times$800 achieve reconstruction quality comparable to those trained at 1200$\times$1200, while requiring less fitting time and fewer primitives.
Therefore, we adopt 800$\times$800 resolution for image-driven editing experiments.

For text-driven editing, we reduce the resolution to 512$\times$512 to avoid out-of-memory issues caused by the high memory footprint of the IP2P model.
However, the rendering resolution after editing can still be higher than 512$\times$512.
}

\begin{table}[htb]
\caption{\hot{Average training and testing PSNR (dB), SSIM, fitting time (min), and number of primitives used for each basic scene of the supernova dataset under different training image resolutions.}}
\vspace{-0.1in}
\centering
\resizebox{\columnwidth}{!}{
\hot{
\begin{tabular}{c|cc|cc|cc}
image  &\multicolumn{2}{c|}{training views} &\multicolumn{2}{c|}{testing views}&fitting&\#\\
resolution   & PSNR$\uparrow$   & SSIM$\uparrow$  & PSNR$\uparrow$   &   SSIM$\uparrow$ &time&primitives\\ \hline
400$\times$400 &27.91 &0.9503&27.37&0.9461&4.24 &37,273 \\
800$\times$800 &29.16&0.9515&29.12&0.9490&6.98 &53,356 \\
1200$\times$1200 &29.74&0.9519&29.34&0.9496&11.28 &61,028 \\
\end{tabular}
}
}
\label{tab:ablation-img-res}
\end{table}

\begin{figure}[htb]
 \begin{center}
 $\begin{array}{c@{\hspace{0.025in}}c@{\hspace{0.025in}}c}
    \includegraphics[height=1.1in]{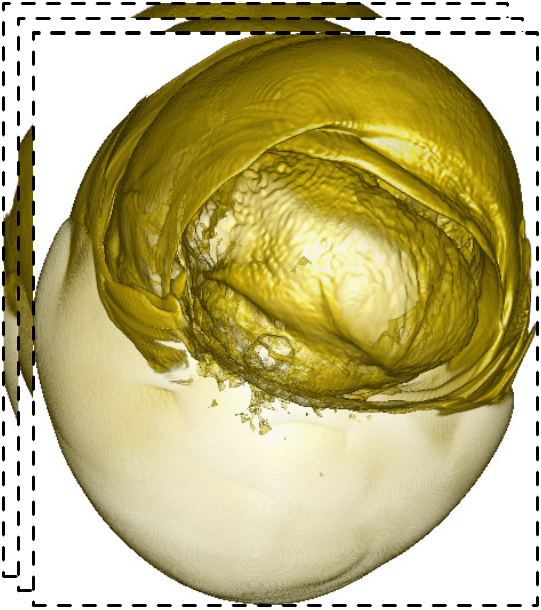}&
    \includegraphics[height=1.1in]{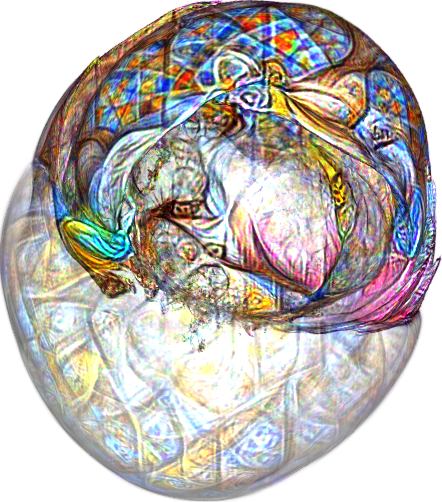}&
    \includegraphics[height=1.1in]{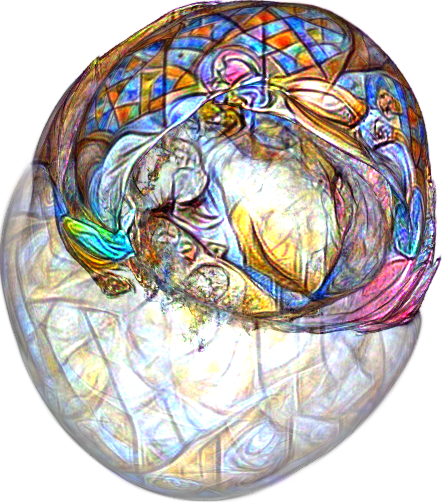}\\
    \mbox{\footnotesize (a) original scene} & \mbox{\footnotesize (b) $\lambda_s=0.0$} & \mbox{\footnotesize (c) $\lambda_s=0.5$} \\
    \includegraphics[height=1.1in]{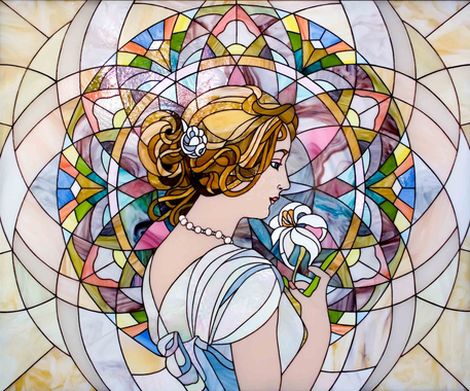}&
    \includegraphics[height=1.1in]{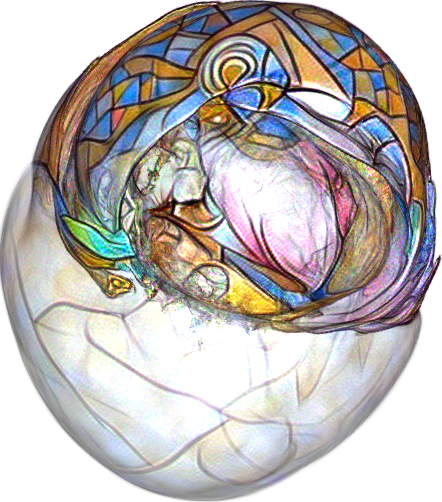}&
    \includegraphics[height=1.1in]{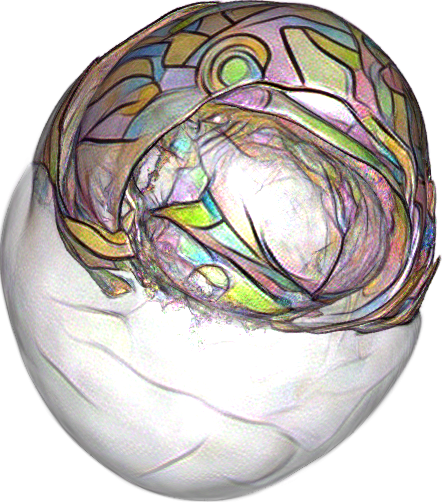}\\
    \mbox{\footnotesize (d) reference style image} & \mbox{\footnotesize (e) $\lambda_s=0.9$} & \mbox{\footnotesize (f) $\lambda_s=1.0$} \\
\end{array}$
\end{center}
\vspace{-.25in} 
\caption{Comparison of TexGS-VolVis image-driven NPSE results under different values of $\lambda_s$.} 
\label{fig:ablation-lambdaS}
\end{figure}

{\bf Evaluation of $\lambda_s$.}
During TexGS-VolVis image-driven NPSE, we set $\lambda_s$ to balance fine stylization details with global style consistency. 
To investigate the impact of this hyperparameter, we optimize TexGS-VolVis using the supernova dataset and a reference style with different $\lambda_s$ values. 
As shown in Figure~\ref{fig:ablation-lambdaS}, the lines in the appearance of the stylized VolVis scene become progressively simplified as $\lambda_s$ increases, reflecting a gradual shift from emphasizing global style patterns to focusing on finer, more localized style details. 
Users can experiment with different $\lambda_s$ values to achieve their desired editing effects.

\vspace{-0.05in}
\section{Additional Results}

\begin{figure}[htb]
 \begin{center}
 $\begin{array}{c@{\hspace{0.025in}}c@{\hspace{0.025in}}c}
    \includegraphics[height=0.955in]{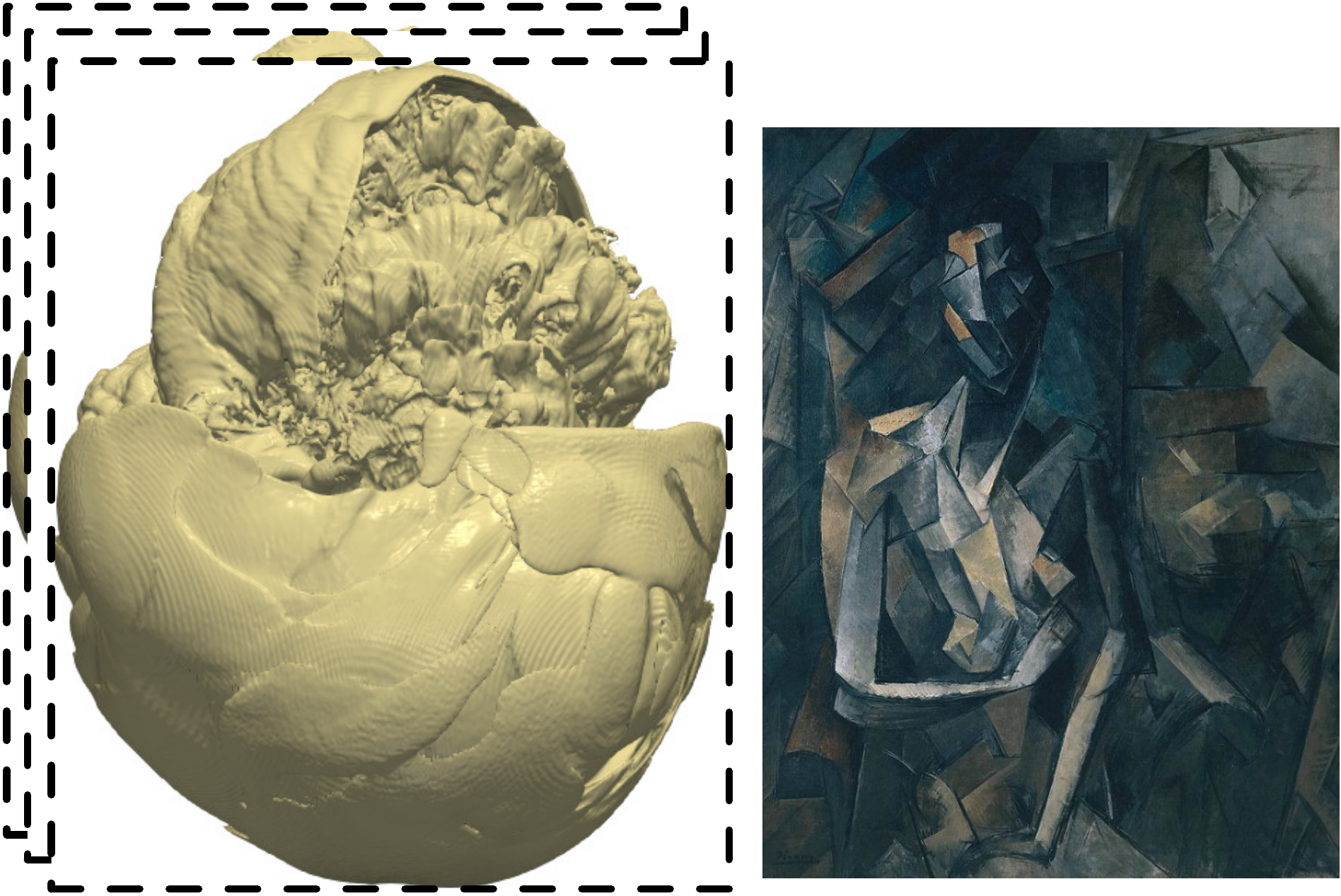}&
    \includegraphics[height=0.955in]{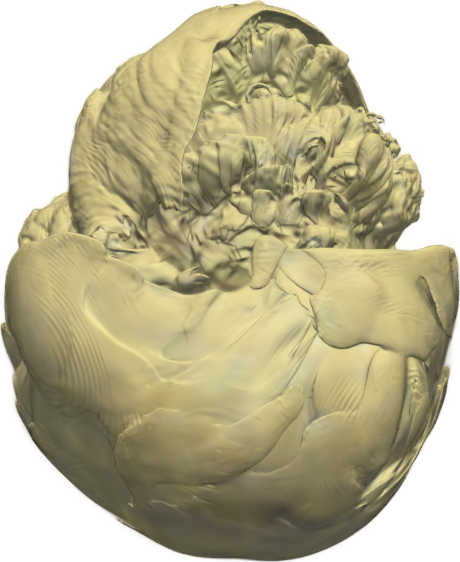}&
    \includegraphics[height=0.955in]{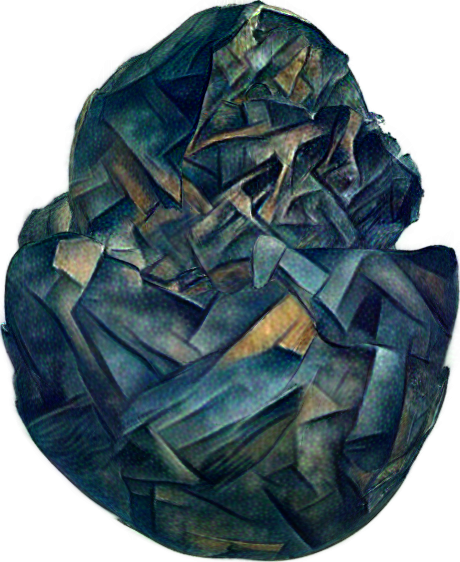}\\
     \includegraphics[height=0.825in]{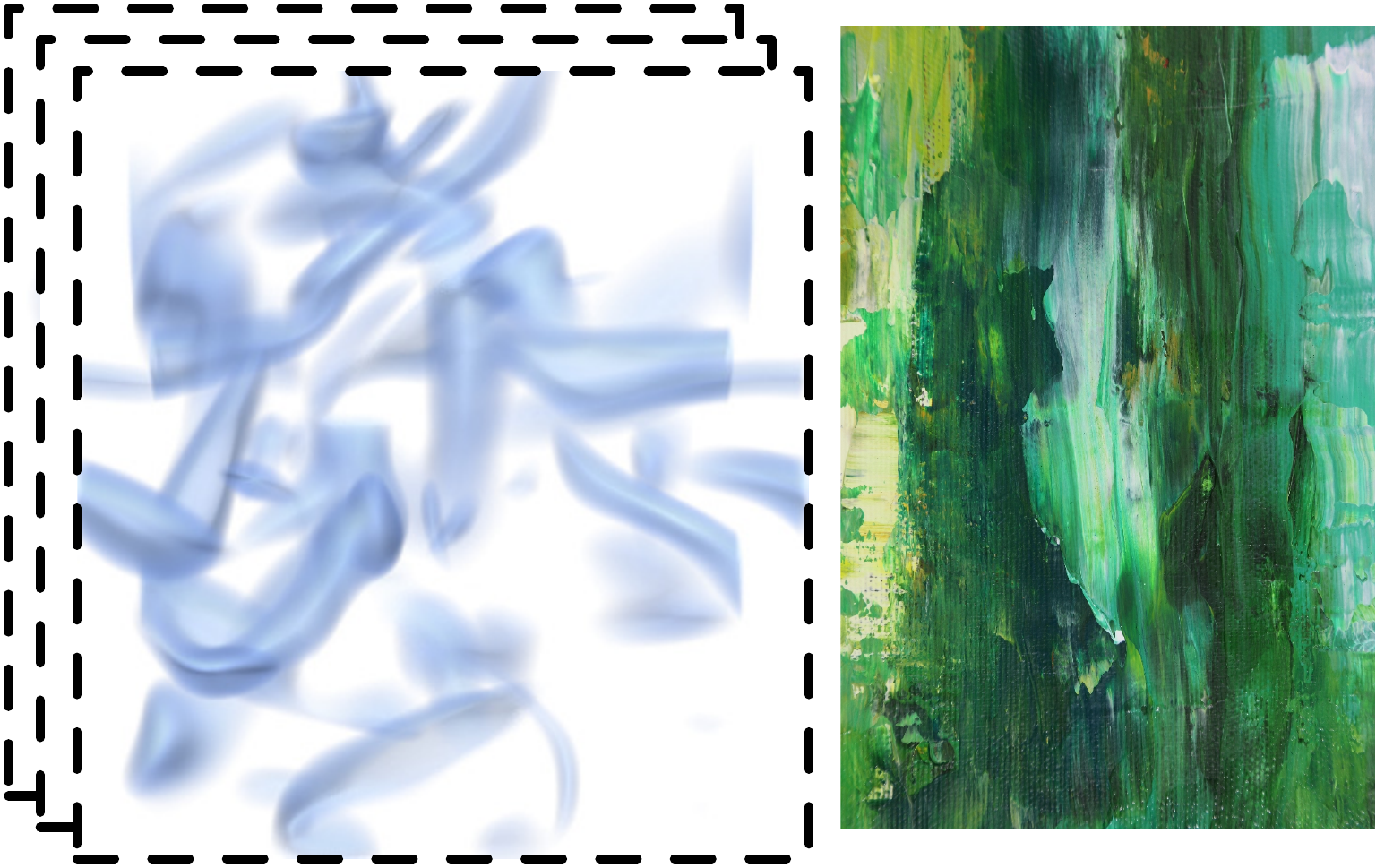}&
    \includegraphics[height=0.825in]{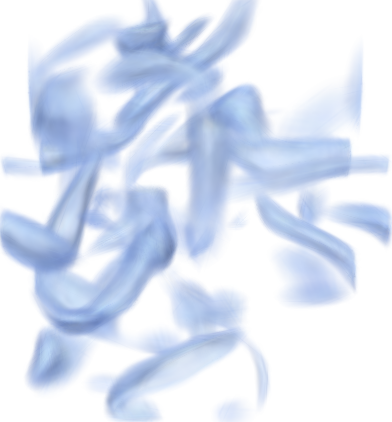}&
    \includegraphics[height=0.825in]{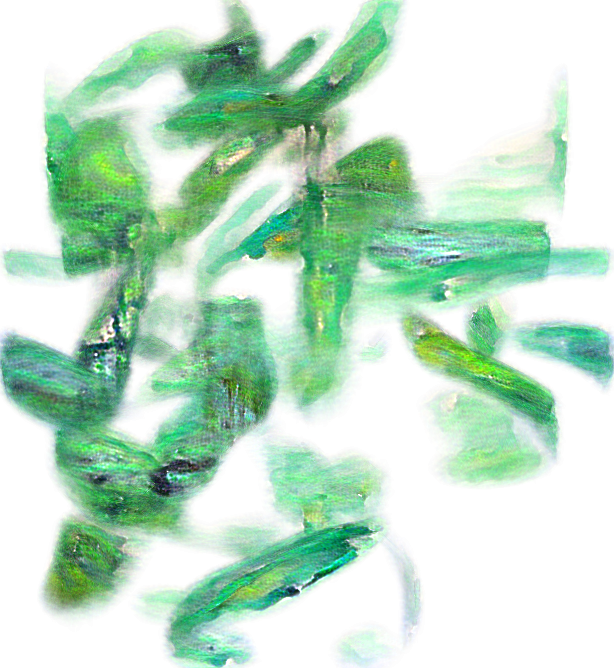}\\
    \includegraphics[height=0.825in]{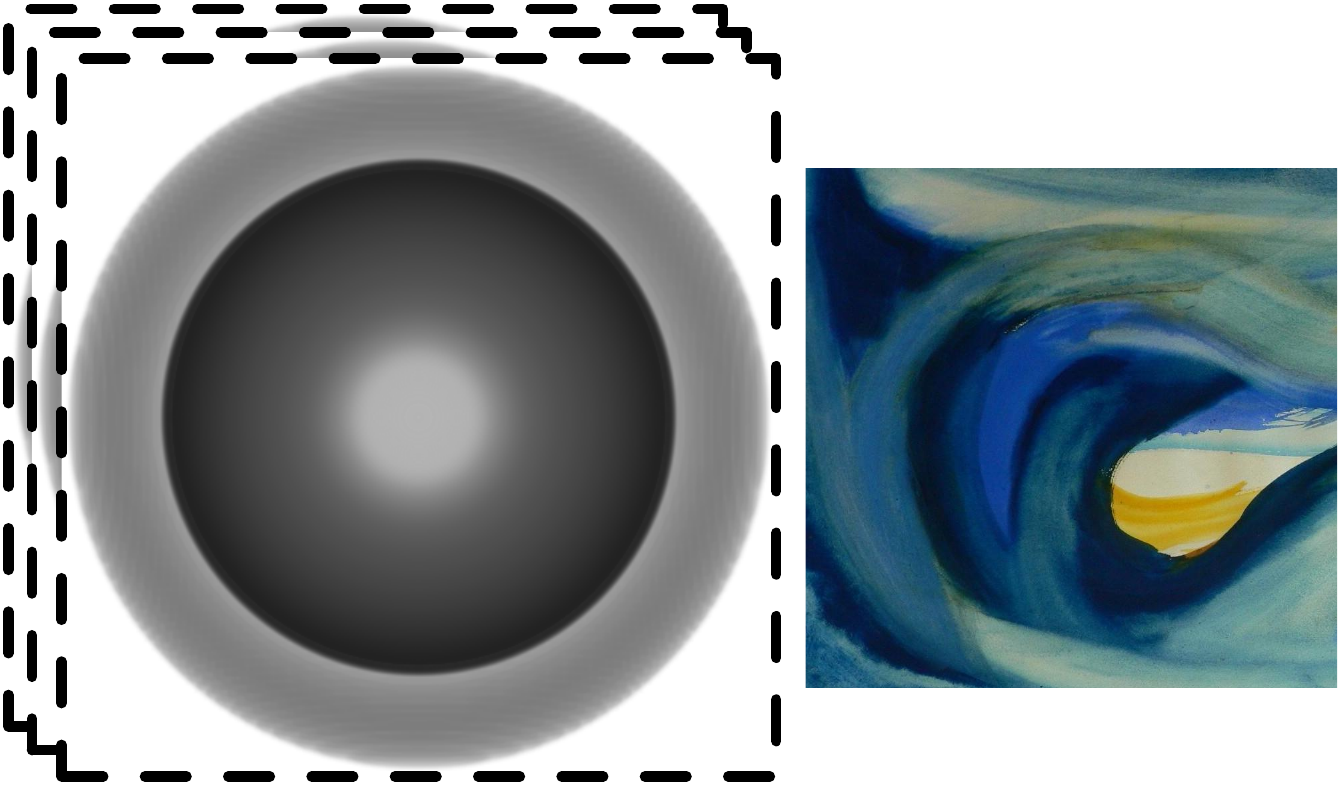}&
    \includegraphics[height=0.825in]{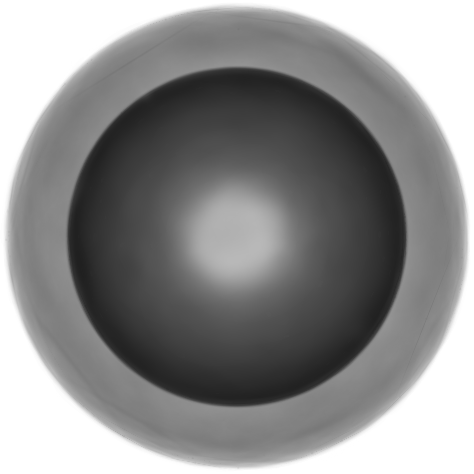}&
    \includegraphics[height=0.825in]{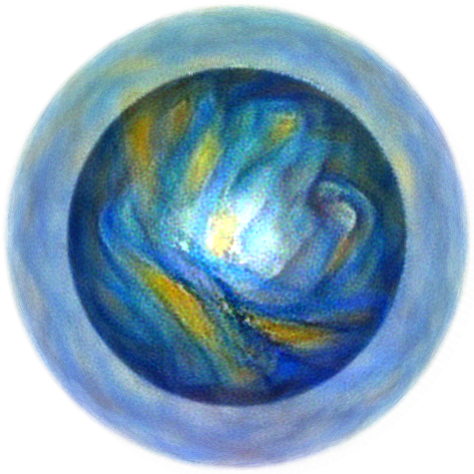}\\
    \mbox{\footnotesize (a) training and style images} & \mbox{\footnotesize (b) scene representation} & \mbox{\footnotesize (c) scene editing} \\
\end{array}$
\end{center}
\vspace{-.25in} 
\caption{\hot{TexGS-VolVis scene representation and editing results. From top to bottom: supernova isosurface, vortex, and synthetic sphere.}} 
\label{fig:flexibility}
\end{figure}

\hot{
{\bf Additional volume datasets.}
To further evaluate the performance of TexGS-VolVis, we tested additional datasets beyond the eight presented in the main paper. Table~\ref{tab:additional-dataset} summarizes these datasets along with their corresponding settings.
Notably, for the fluid simulation dataset vortex~\cite{silver1997tracking}, we rendered scenes without using the NVIDIA IndeX plugin in ParaView. 
While this choice increased rendering time, it allowed us to obtain volumetric scenes lacking clearly defined surfaces, providing a valuable test case to assess the robustness of our method under more challenging visualization conditions.

\begin{table}[htb]
\caption{\hot{Additional datasets and their settings used in the appendix. The rendering time is for DVR using ParaView.}}
\vspace{-0.1in}
\centering
\hot{
\resizebox{\columnwidth}{!}{
\begin{tabular}{c|cc|ccc}
   	     & volume     &image  		&\# basic &  volume  & rendering     \\ 
 dataset & resolution &resolution   &  scenes &  size (MB)  &  time (ms) \\ \hline
 combustion &480$\times$720$\times$120 &800$\times$800 &5 &158  &29.96 \\
 rotstrat & 2048$\times$2048$\times$2048 &800$\times$800 &1 &32,768   & 1649.52 \\
 sphere  &512$\times$512$\times$512 &800$\times$800 &1 & 512  & 78.62\\
 vortex &1024$\times$1024$\times$1024 &800$\times$800 &1 &4,096  &577.5  \\

\end{tabular}
}}
\label{tab:additional-dataset}
\end{table}

{\bf Flexibility of scene representation and editing.}
As shown in Figure~\ref{fig:flexibility}, beyond the datasets included in the paper, our representation and editing method also generalizes to other forms of VolVis scenes.
It supports fully solid isosurface scenes (e.g., supernova) as well as``faded'' volumes without clearly defined surfaces (e.g., vortex).
We also evaluated our method on a synthetic sphere dataset, where the center has high opacity and gradually fades outward, forming a low-opacity shell.
Our results show that TexGS-VolVis can successfully stylize such cases, demonstrating its robustness even in sparsely defined volumetric regions.
}

\begin{figure}[htb]
 \begin{center}
\includegraphics[width=1.0\linewidth]{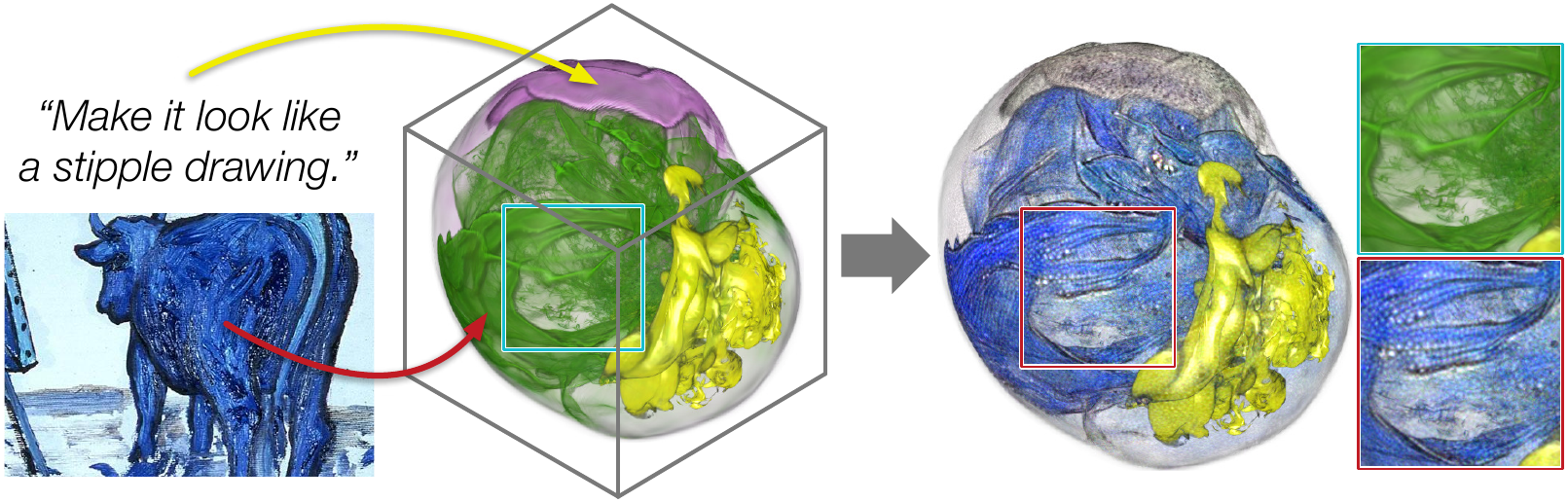}\\
\mbox{\footnotesize (a) blurry stylization}\\
 $\begin{array}{c@{\hspace{0.05in}}c}
 \includegraphics[height=0.68in]{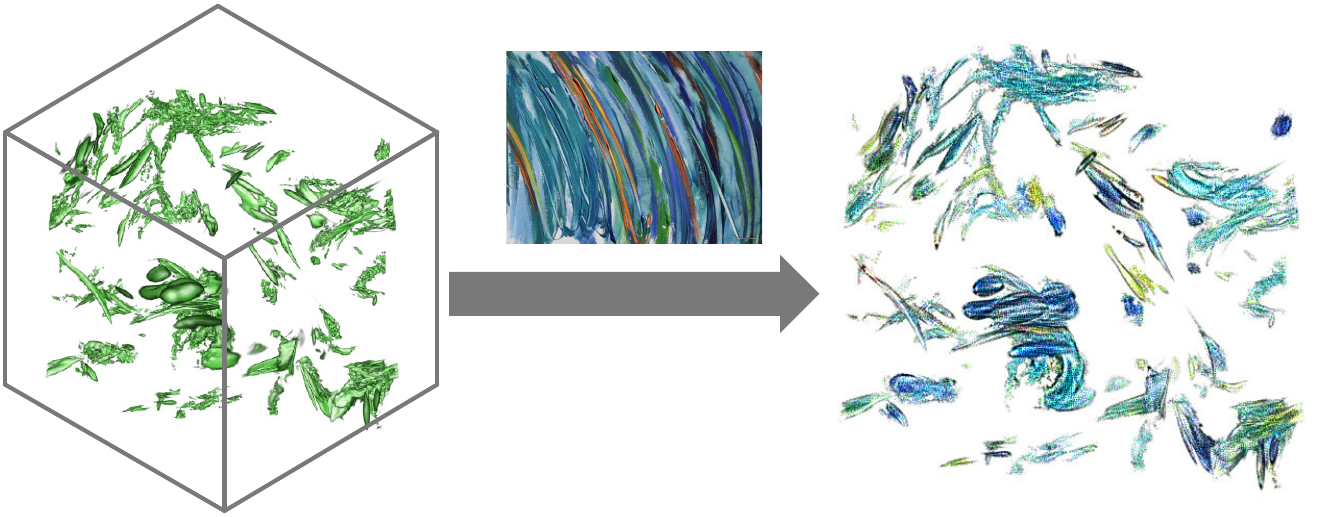}&
\includegraphics[height=0.68in]{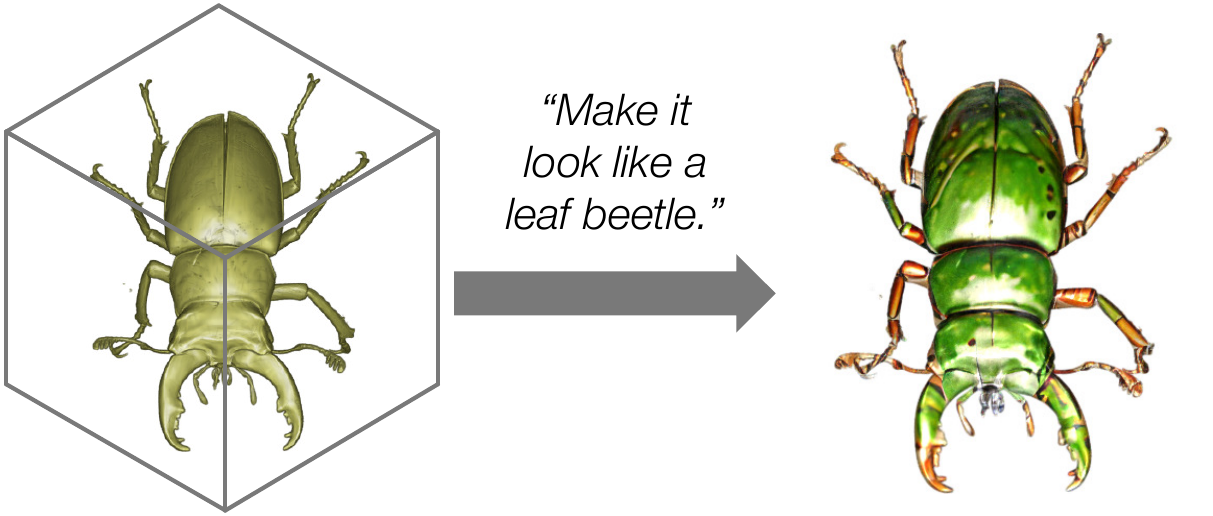}\\
\mbox{\footnotesize (b) complex structure} & 
\mbox{\footnotesize (c) lexical misinterpretation}\\
\end{array}$
\end{center}
\vspace{-.25in} 
\caption{\hot{Failure cases of TexGS-VolVis for image-driven and text-driven scene editing. (a) to (c): supernova, rotstrat, and beetle.}} 
\label{fig:failure-cases}
\end{figure}

\hot{
{\bf Failure editing cases.}
While TexGS-VolVis performs well across a range of VolVis scenes, we observe several scenarios where the model may produce failure editing results.
Figure~\ref{fig:failure-cases} shows the three failure cases we found.
First, flat or low-texture styles tend to produce overly smoothed stylization results. As shown in the zoomed-in region of Figure~\ref{fig:failure-cases}~(a), the rich internal details of the original scene become visually indistinct after stylization.
Second, TexGS-VolVis performs poorly on scenes with fragmented and complex structures, as shown in Figure~\ref{fig:failure-cases}~(b).
The lack of large, continuous surfaces makes it difficult to apply the style coherently.
Third, our text-driven editing relies heavily on the pretrained IP2P model. 
When the model misinterprets the prompt, it can lead to unexpected results.
As shown in Figure~\ref{fig:failure-cases}~(c), given the prompt ``{\em Make it look like a leaf beetle},'' IP2P incorrectly associates the beetle with a green appearance, whereas real leaf beetles often have red, yellow, or black coloration.
}

\begin{figure}[htb]
 \begin{center}
 \includegraphics[width=1.0\linewidth]{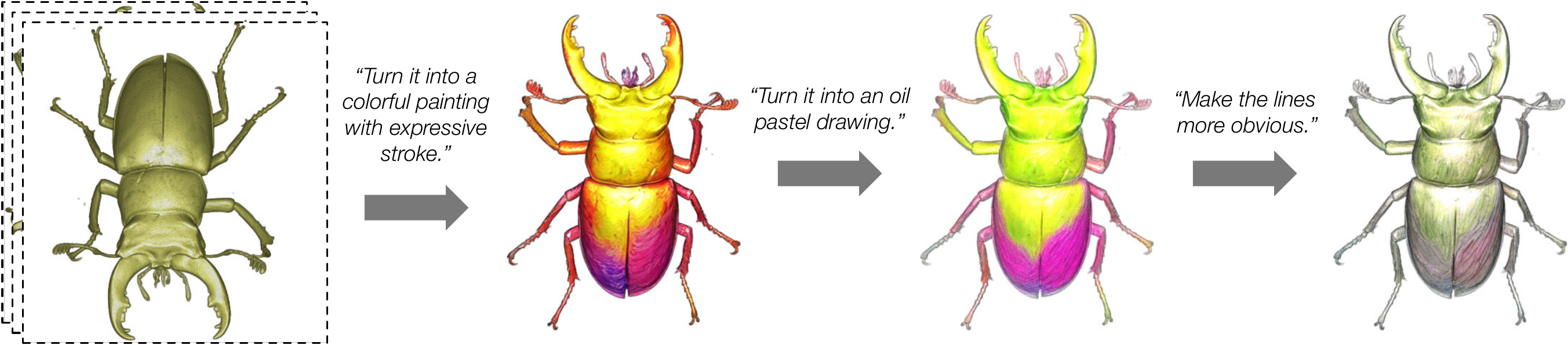}
\end{center}
\vspace{-.2in} 
\caption{TexGS-VolVis editing results via text prompt refinement using the beetle dataset.} 
\label{fig:text-prompt-refinement}
\end{figure}

{\bf Text prompt refinement.}
In addition to editing the original VolVis scene with a text prompt, TexGS-VolVis enables text-driven NPSE on already stylized VolVis scenes, allowing for further refinements using additional text prompts.
We refer to this iterative process as {\em text prompt refinement}.
Figure~\ref{fig:text-prompt-refinement} illustrates such an example using the beetle dataset.
When a new text prompt is applied to an already stylized scene, the previous style pattern is partially preserved while being updated according to the latest prompt.
This iterative approach offers users greater flexibility, allowing them to refine their edits step by step to achieve the desired effects.

\begin{figure}[htb]
 \begin{center}
 \includegraphics[width=1.0\linewidth]{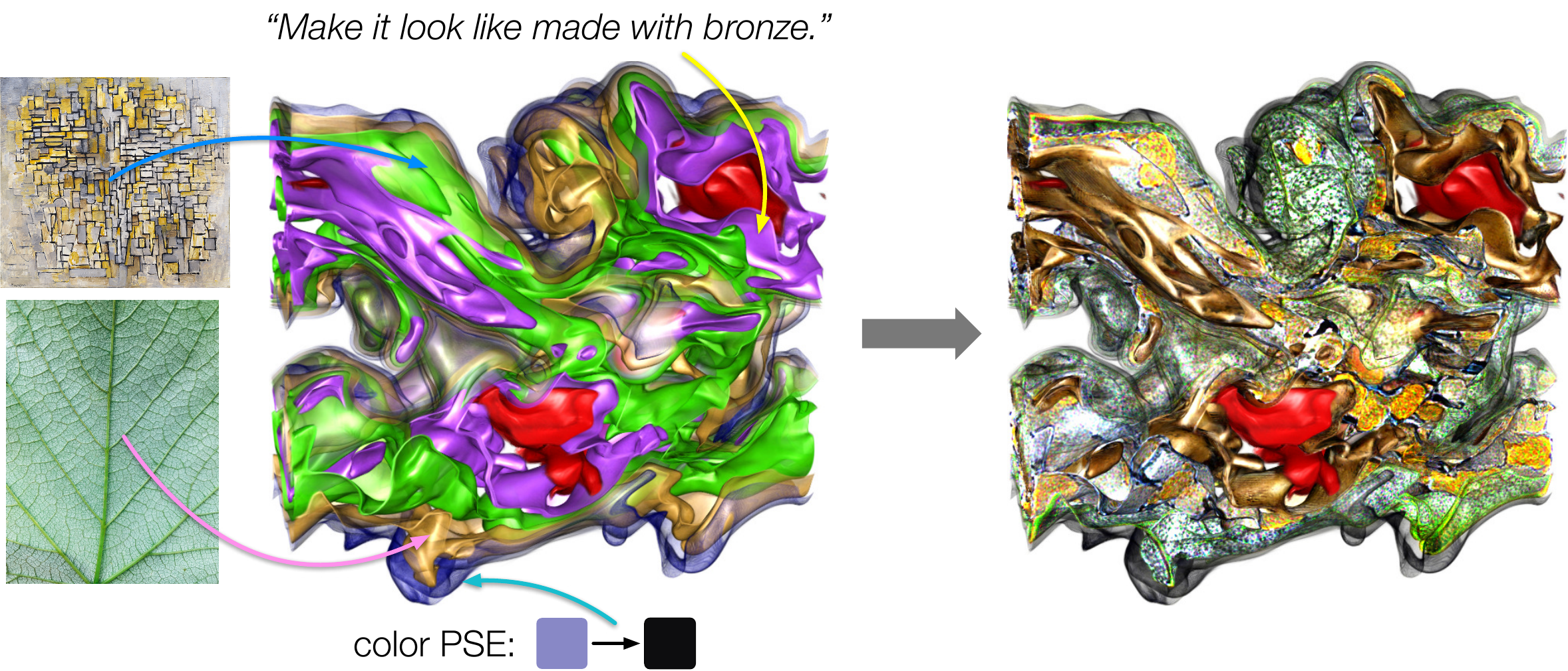}
\end{center}
\vspace{-.2in} 
\caption{TexGS-VolVis's NPSE and PSE results on the combustion dataset with five basic scenes.} 
\label{fig:combustion}
\end{figure}

\begin{figure*}[htb]
 \begin{center}
 $\begin{array}{c@{\hspace{0.05in}}c@{\hspace{0.05in}}c@{\hspace{0.05in}}c@{\hspace{0.05in}}c@{\hspace{0.05in}}c}
 \includegraphics[width=0.15\linewidth]{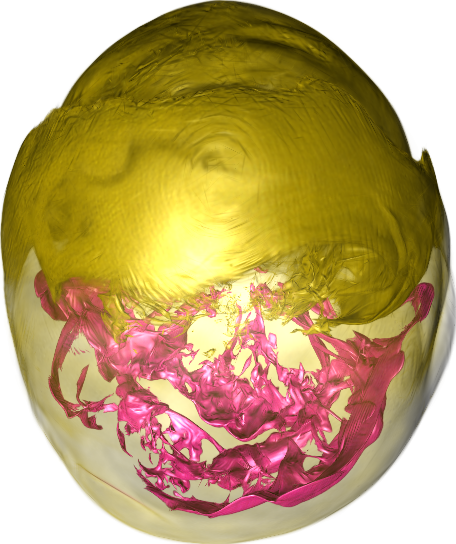}&
 \includegraphics[width=0.15\linewidth]{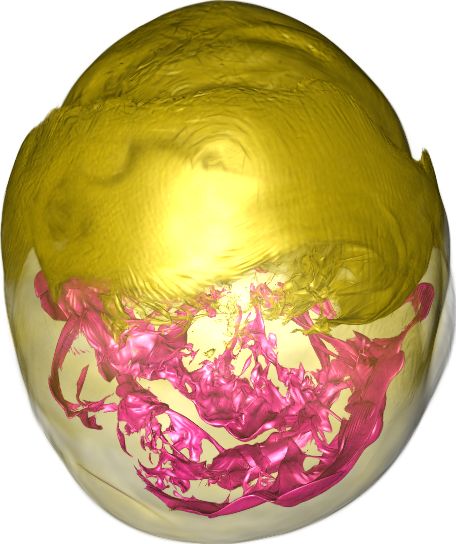}&
 \includegraphics[width=0.15\linewidth]{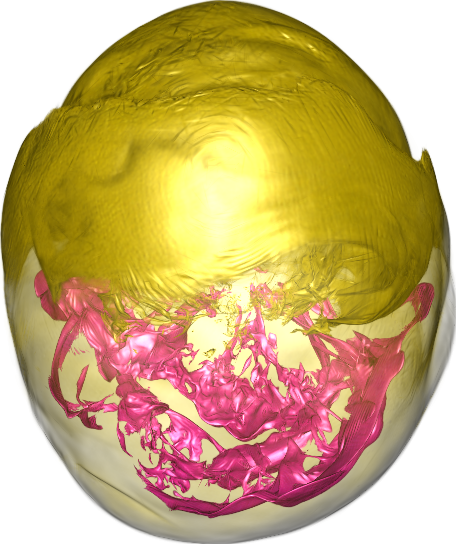}&
 \includegraphics[width=0.15\linewidth]{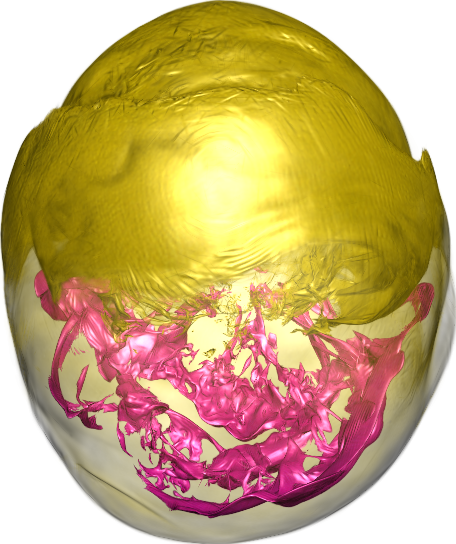}&
 \includegraphics[width=0.15\linewidth]{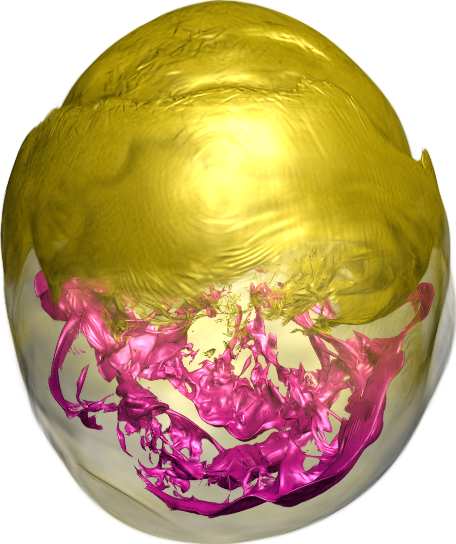}&
\includegraphics[width=0.15\linewidth]{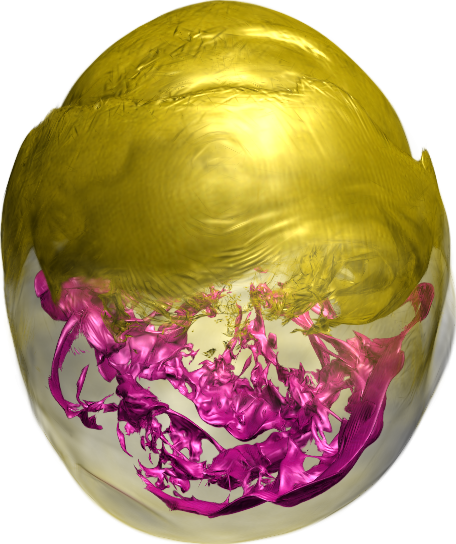}\\
\mbox{\footnotesize $(-90^\circ, -30^\circ)$} & 
\mbox{\footnotesize $(-90^\circ, -15^\circ)$} & 
\mbox{\footnotesize $(-90^\circ, 0^\circ)$} &
\mbox{\footnotesize $(-90^\circ, 15^\circ)$} &
\mbox{\footnotesize $(-90^\circ, 30^\circ)$} & 
\mbox{\footnotesize $(-90^\circ, 45^\circ)$}\\
 \includegraphics[width=0.15\linewidth]{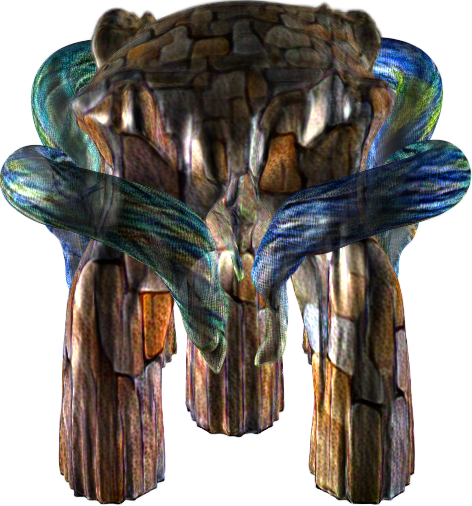}&
  \includegraphics[width=0.15\linewidth]{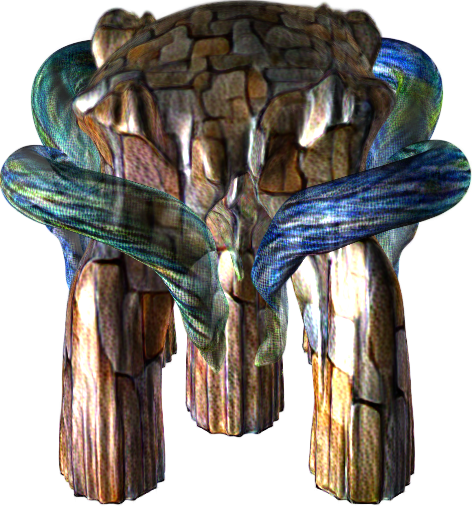}&
  \includegraphics[width=0.15\linewidth]{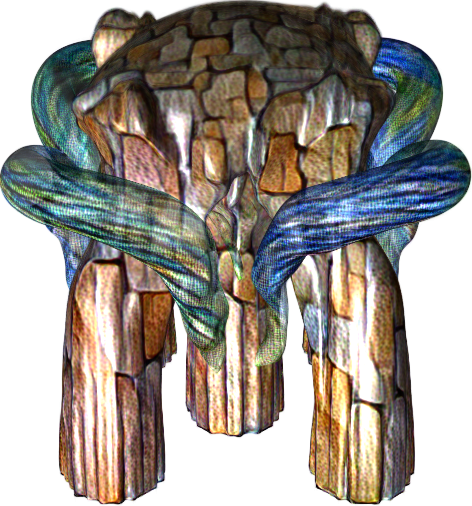}&
 \includegraphics[width=0.15\linewidth]{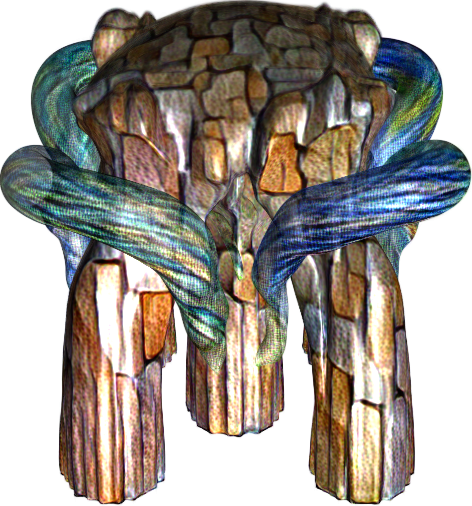}&
 \includegraphics[width=0.15\linewidth]{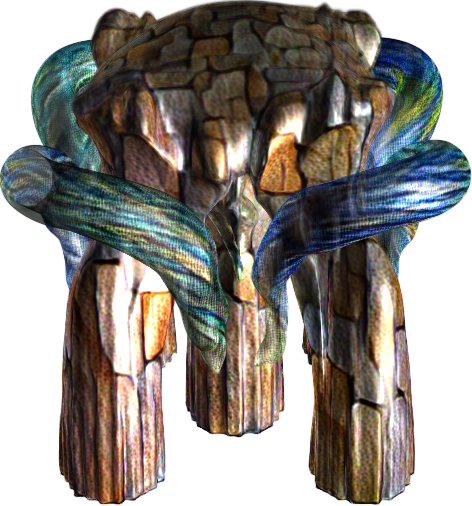}&
\includegraphics[width=0.15\linewidth]{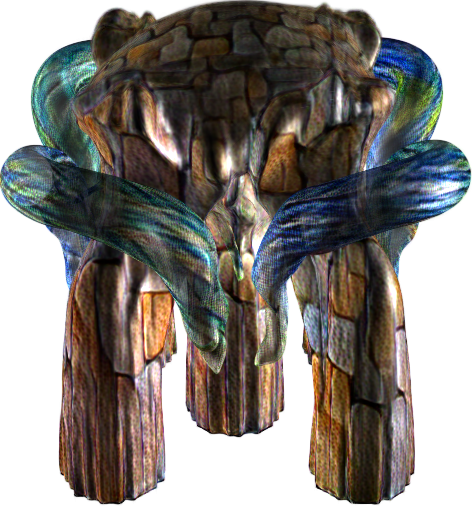}\\
\mbox{\footnotesize $(-180^\circ, 0^\circ)$} & 
\mbox{\footnotesize $(-144^\circ, 0^\circ)$} & 
\mbox{\footnotesize $(-108^\circ, 0^\circ)$} &   
\mbox{\footnotesize $(-72^\circ, 0^\circ)$} & 
\mbox{\footnotesize $(-36^\circ, 0^\circ)$} & 
\mbox{\footnotesize $(0^\circ, 0^\circ)$}\\
\end{array}$
\end{center}
\vspace{-.25in} 
\caption{TexGS-VolVis relighting results on the supernova and five jet datasets. The light direction is expressed as azimuthal and polar angles.} 
\label{fig:lighting-dir-change}
\end{figure*}

{\bf VolVis scene with multiple basic TFs.}
In this paper, we limit the VolVis scenes to a maximum of two basic scenes to clearly showcase the stylization results for each.
However, our framework is not restricted to this number and can accommodate more basic scenes.
Figure~\ref{fig:combustion} illustrates the editing results of TexGS-VolVis on the combustion dataset, which contains five basic scenes.
TexGS-VolVis allows independent editing of each basic scene and seamlessly composes them into a complete VolVis scene.
Since this composition does not require additional optimization, TexGS-VolVis can efficiently handle volumetric datasets with multiple basic scenes without incurring extra training costs.

{\bf Light direction change.}
To better demonstrate TexGS-VolVis's performance in handling light direction changes, we present the relighting results for two datasets: supernova and stylized five jets.
As shown in Figure~\ref{fig:lighting-dir-change}, TexGS-VolVis enables smooth and consistent relighting when adjusting the lighting direction using azimuthal and polar angles.

\begin{figure}[htb]
\begin{center}
	\includegraphics[width=1.0\linewidth]{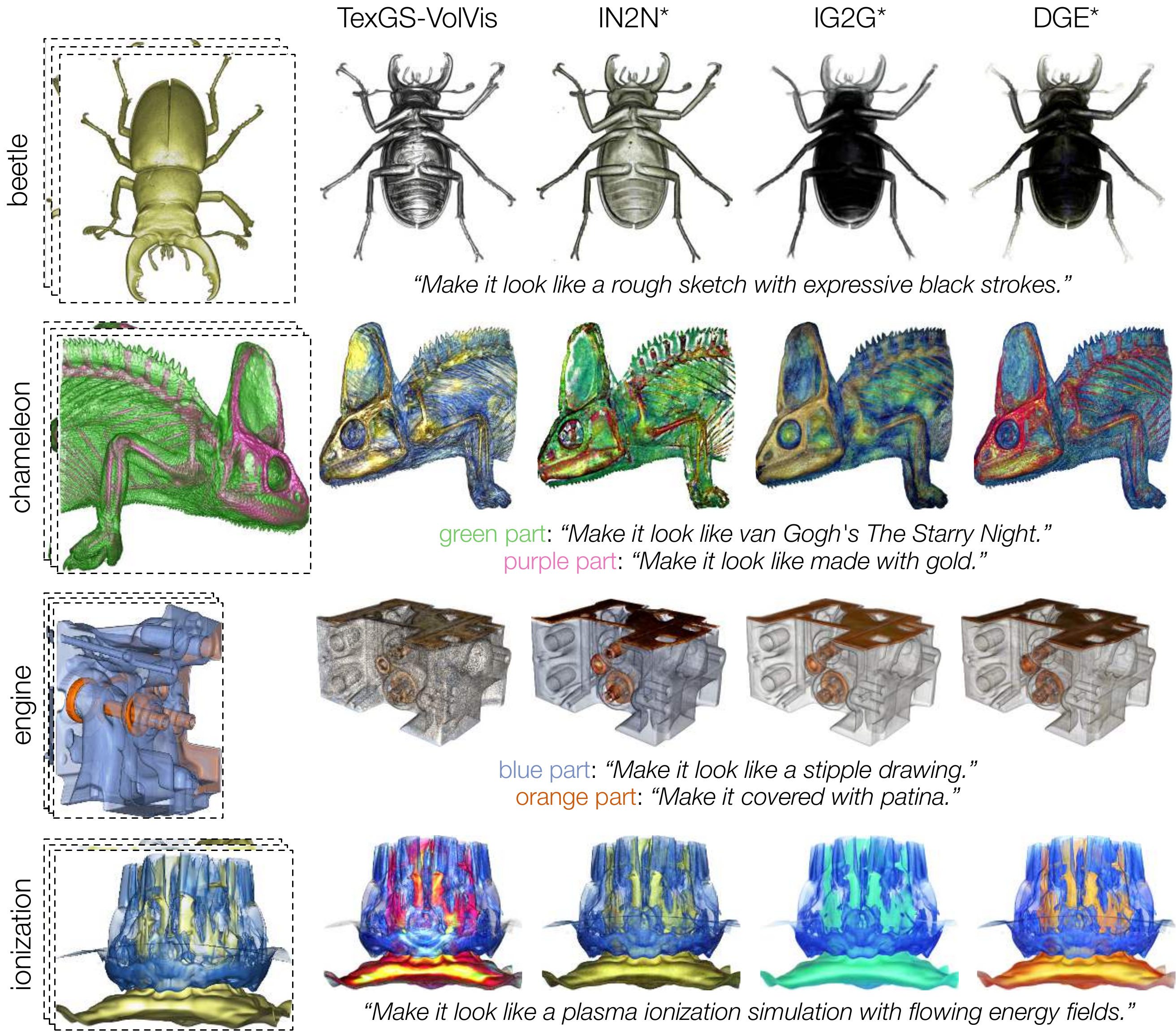}
	\end{center}
	\vspace{-.25in}
\caption{Comparison of TexGS-VolVis with geometry-frozen baseline methods (marked with $\star$) for text-driven NPSE across the beetle, chameleon, engine, and ionization datasets.} 
\label{fig:baseline-text-NPSE-geo}
\end{figure}

{\bf Text-driven NPSE baselines without geometry update.}
For TexGS-VolVis, the simplest approach to ensure geometry-consistent text-driven editing is to freeze the geometry representation within the scene model while performing NPSE. 
However, this constraint reduces the model's expressiveness for explicit methods like GS, as geometry and appearance representations are inherently coupled. 
Figure~\ref{fig:baseline-text-NPSE-geo} presents the NPSE results of TexGS-VolVis and baseline methods with geometry-related parameters frozen during text-driven editing. 
While this eliminates the elliptical artifacts in IG2G and DGE, the overall stylization quality of these baseline methods remains unsatisfactory due to the diminished expressive capability in appearance representation. 
In contrast, TexGS-VolVis leverages the texture attribute to decouple geometry from appearance modeling, enabling geometry-consistent editing while preserving rich visual expressiveness.

\vspace{-0.05in}
\section{Additional Discussion and Details}

\hot{
{\bf Model size breakdown.}
Table~\ref{tab:memory-break-down} shows the model size breakdown when representing different datasets with varying numbers of primitives.
All attributes are stored as 32-bit floating-point values. 
For all datasets, the total number of texels $T_\text{total}$ is fixed at $1\times10^7$, resulting in approximately 114.5 MB of memory usage for the texture attribute per basic scene, regardless of the number of primitives used.
Note that the spherical harmonic parameters and view-independent color $\mathbf{c}_\text{ind}$ attribute are only used for early phases optimization, which are not retrained in the final scene representation.
The texture attribute typically accounts for over 95\% of the total model size, highlighting the necessity of exploring texture compression in future work.

Except for the texture attribute, each splat contains geometry attributes $\{\bm{\mu}, \mathbf{t}_u, \mathbf{t}_v, s_u, s_v, o \}$ and shading attributes $\{k_a, k_d, k_s, \beta\}$.
In practice, we use a quaternion parameter to represent $\{\mathbf{t}_u, \mathbf{t}_v\}$ of a primitive.
Therefore, given the number of primitives $N$, the number of parameters of a TexGS-VolVis model for the geometry and shading attributes can be calculated as
\begin{equation}
\begin{split}
 \text{\# params }&= \underbrace{3N+4N+N+N+N}_{\text{\# params for geometry attributes}} \\
  &+ \underbrace{N+N+N+N.}_{\text{\# params for shading attributes}}
\end{split}
\end{equation}

\begin{table}[htb]
\caption{\hot{TexGS-VolVis model size (MB) breakdown for representing different datasets.}}
\vspace{-0.1in}
\centering
\resizebox{3in}{!}{
\hot{
\begin{tabular}{c|c|ccc|c}
 & \# basic   & geometry   & shading    & texture   & total \# \\ 
dataset &  scenes  & attributes   & attributes   &  attribute  &  primitives\\ \hline
beetle &1 &1.8 &0.7 &114.5 &47,767\\
ionization &2 &4.6 &1.8 &228.9 &119,396  \\
wood&1 &5.3 &2.1 &114.5 &138,087  \\
mantle &2 &6.9 &2.7 &228.9 &179,688  \\ 
\end{tabular}
}
}
\label{tab:memory-break-down}
\end{table}

{\bf TexGS-VolVis fitting performance.}
In our workflow, the scene representation produced by TexGS-VolVis forms the foundation for the subsequent scene editing. 
We report quantitative model fitting results in Table~\ref{tab:model-fitting-res}, evaluated across the eight datasets presented in Table 1 of the paper.
After three optimization phases (2DGS, shading attributes, and the texture attribute), our method achieves acceptable PSNR and SSIM reconstruction accuracy within a few minutes on all these datasets. 
Moreover, the training process requires relatively low CPU and GPU memory, which is related to the training image resolution.

By comparing reconstruction accuracy across different datasets, we observe that TexGS-VolVis finds it more challenging to fit VolVis scenes with TFs that produce rugged or highly detailed surfaces (e.g., wood) instead of smoother ones (e.g., five jets).
This is because rugged surfaces tend to produce more complex lighting effects in the multi-view images, making it more difficult for the model to reconstruct the correct underlying geometry from the observations. 

Among the three optimization phases, optimizing the shading attributes typically improves reconstruction accuracy. 
This is mainly due to the shading attributes' ability to model view-dependent color variations.
In contrast, optimizing the texture attribute may slightly reduce reconstruction accuracy, as it introduces an additional sparsity regularization term (refer to Section 3.4 in the paper) that encourages the texture values to remain minimal.
}

\begin{table}[htb]
	\caption{\hot{Average PSNR (dB), SSIM, fitting time (min), CPU/GPU memory (GB), and number of primitives for the eight datasets shown in Table 1 of the paper.}}
	\vspace{-0.1in}
	\centering
	\hot{
	\resizebox{\columnwidth}{!}{
		\begin{tabular}{c|c|ccccc}
			 &  &                              &                             & fitting  & CPU/GPU & \# \\
	    dataset & fitting phase & PSNR$\uparrow$ & SSIM$\uparrow$ & time         & memory & primitives \\
			\hline 
			               	& 2DGS      &29.37 &0.9776&1.18&10.1/5.1 & 61,495\\
 			five jets		& + shading attributes  &36.26&0.9906&5.08&10.1/5.1 &56,836        \\
							& + texture attribute &36.83&0.9889&0.8&10.0/5.6&56,836 \\
			\hdashline 
		               		& 2DGS      &27.79&0.9518&1.17&10.1/5.1 &83,453\\
 			mantle			& + shading attributes  &31.12&0.9735&5.69&10.1/5.1&89,844       \\
							& + texture attribute &30.46&0.9708&0.46&10.1/5.5 &89,844  \\
			\hdashline 
		               		& 2DGS      &27.64&0.9356&1.14&9.9/5.1 &48,145\\
 			supernova			& + shading attributes  &29.34&0.9508&4.96&10.0/5.1 &53,356      \\
							& + texture attribute &29.12&0.9490&0.88&10.1/5.6 &53,356  \\
			\hdashline
		               		& 2DGS      &23.87&0.8243&1.52&10.0/5.1 &104,409\\
 			wood			& + shading attributes  &24.60&0.8433&6.45&10.0/5.2 &138,087      \\
							& + texture attribute &24.58&0.8565&0.9&10.1/5.6 &138,087 \\
			\hline 
		               		& 2DGS      &33.50&0.9735&0.80&4.8/2.1 &54,796\\
 			beetle			& + shading attributes  &33.25&0.9791&3.38&4.9/2.1 &47,767    \\
							& + texture attribute &33.37&0.9811&0.72&4.8/2.6 &47,767 \\
			\hdashline 
		               		& 2DGS      &25.85&0.9203&0.81&4.9/2.1 &71,186\\
 			chameleon		& + shading attributes  &26.95&0.9377&3.78&4.9/2.1&82,081      \\
							& + texture attribute &26.48&0.9408&0.78&4.8/2.6&82,081  \\
			\hdashline
		               		& 2DGS      &30.15&0.9585&0.85&4.8/2.2 &50,628\\
 			engine			& + shading attributes  &33.25&0.9762&4.06&4.9/2.1 &49,151      \\
							& + texture attribute &33.29&0.9754&0.76&4.8/2.6 &49,151 \\
			\hdashline 
		               		& 2DGS      &29.35&0.9568&0.84&4.9/2.1 &56,555\\
 			ionization		& + shading attributes  &32.12&0.9722&3.96&4.9/2.1   &59,698    \\
							& + texture attribute &31.64&0.9712&0.75&4.8/2.6 &59,698 \\

		\end{tabular}
	}}
	\label{tab:model-fitting-res}
\end{table}

\begin{figure}[htb]
 \begin{center}
 $\begin{array}{c@{\hspace{0.025in}}c@{\hspace{0.025in}}c}
 \includegraphics[width=0.315\linewidth]{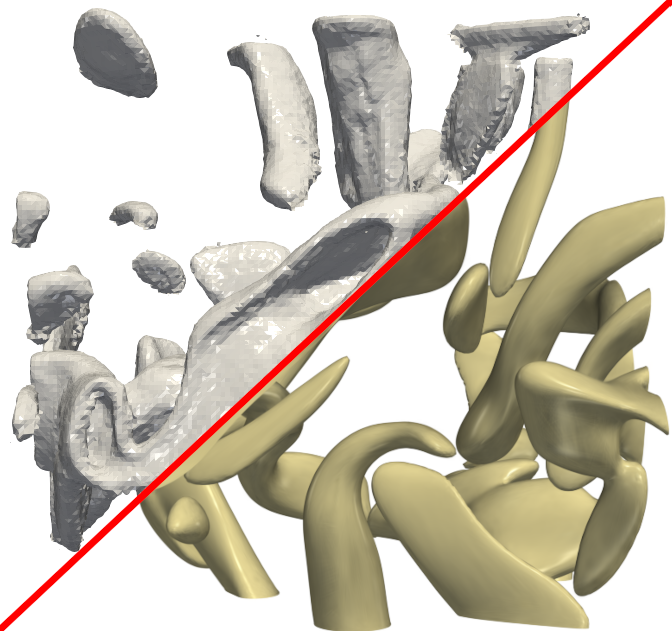}&
 \includegraphics[width=0.315\linewidth]{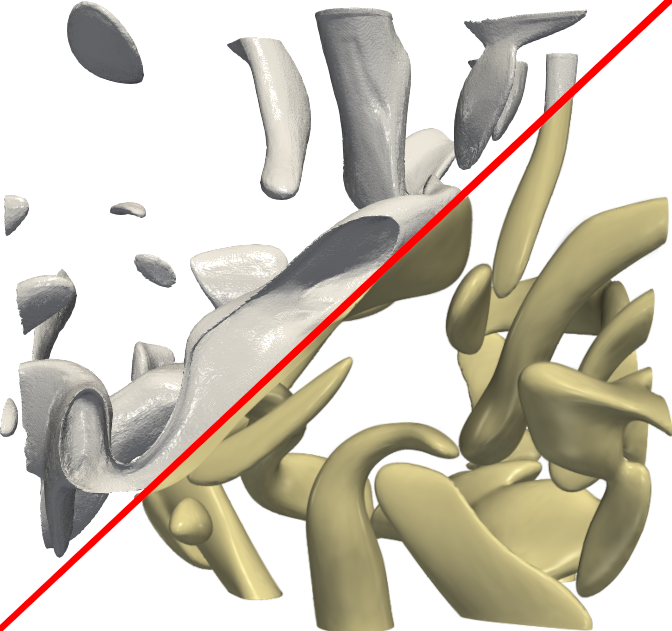}&
 \includegraphics[width=0.315\linewidth]{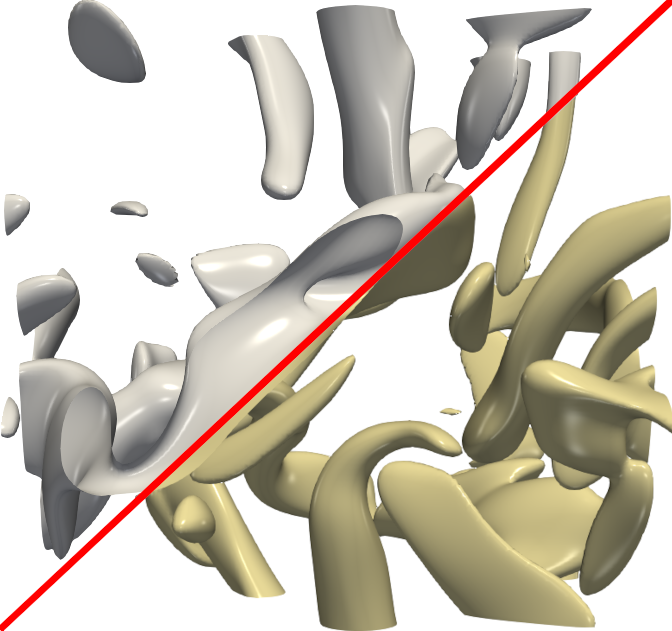}\\
\mbox{\footnotesize (a) 3DGS} & \mbox{\footnotesize (b) 2DGS} & \mbox{\footnotesize (c) ground-truth} 
\end{array}$
\end{center}
\vspace{-.25in} 
\caption{Extracted mesh (top-left) and NVS results (bottom-right) for the vortex dataset. While both methods achieve comparable NVS accuracy, 2DGS significantly outperforms 3DGS in geometry reconstruction.} 
\label{fig:geometry-2DGS-vs-3DGS}
\end{figure}

\hot{
{\bf 3DGS vs.\ 2DGS.}
Existing VolVis scene representation methods~\cite{Niedermayr-arxiv24, Tang-PVIS25} are primarily based on 3DGS, whereas TexGS-VolVis opts for 2DGS.
Unlike 3DGS, 2DGS employs explicit ray-splat intersection, ensuring perspective-correct splatting and yielding more accurate geometry reconstruction.
Figure~\ref{fig:geometry-2DGS-vs-3DGS} compares extracted meshes from 3DGS and 2DGS using truncated signed distance fusion~\cite{Newcombe-TSDF}, along with their respective NVS results. 
These comparisons demonstrate that 2DGS provides a more precise geometric representation than 3DGS.
Furthermore, 2DGS inherently models surface normals and is naturally compatible with 2D texture maps, making it an ideal representation for VolVis scenes, especially when aiming for geometry-consistent NPSE.

Given that 2DGS provides a more precise geometry representation than 3DGS, an intuitive editing strategy would be to directly perform edits on the surface extracted from 2DGS.
This approach is similar to the one proposed in Texture-GS~\cite{Xu-ECCV24}.
However, as noted in Texture-GS's implementation, 
such editing only works well for objects with simple geometry, which is not the case for most VolVis scenes.
Moreover, there remains a gap in accuracy between the isosurfaces extracted from 2DGS and the ground truth geometry, which poses challenges for achieving high-quality stylization.
}

{\bf Participant voting on each dataset.}
Figure 5 in the paper presents the aggregated participant voting results from the user study, while Table~\ref{tab:user-study-votes} provides a detailed breakdown of votes for each dataset. 
The results show that the majority of votes for TexGS-VolVis fall into the ``best'' and ``second-best'' categories, highlighting the effectiveness of our method. 
The ``worst'' votes are primarily concentrated in the five jets and engine datasets, likely because some participants preferred low-frequency style patterns over highly detailed, high-frequency ones.

\begin{table}[htb]
	\caption{The votes from ten participants ranking the stylization outcomes of image-driven and text-driven NPSE (Figures 2 and 3 in the paper).}
	\vspace{-0.1in}
	\centering
	\resizebox{3in}{!}{
		\begin{tabular}{c|c|cccc}
			  &    & \multicolumn{4}{c}{\# votes} \\
			dataset & method & best & second-best & second-worst & worst \\
			\hline 
			               	& ARF      &0  &3   &4 &3\\
 			five jets		& StyleRF-VolVis  &1 &2  &4    &3       \\
							& StyleSplat &3  &3   &2  &2  \\
			  				& TexGS-VolVis 	   &6  &2   &0 &2   \\
			\hdashline 
			               	& ARF      &0  &3   &3 &4\\
 			wood			& StyleRF-VolVis  &0 &3  &7    &0       \\
							& StyleSplat &2  &2   &0  &6  \\
			  				& TexGS-VolVis 	   &8  &2   &0 &0   \\
			\hdashline
			               	& ARF      &0&2&2&6\\
 			supernova  		& StyleRF-VolVis  &1&4&4&1       \\
							& StyleSplat &0&3&4&3  \\
			  				& TexGS-VolVis 	   &9&1&0&0   \\
			\hdashline 
			               	& ARF     &0&0&1&9\\
 			mantle 			& StyleRF-VolVis  &3&6&1&0      \\
							& StyleSplat &3&1&5&1  \\
			  				& TexGS-VolVis 	   &4&3&3&0  \\
			\hline 
			               	& IN2N      &1&0&1&8\\
 			chameleon		& IG2G  &0&7&2&1      \\
							& DGE &0&3&6&1  \\
			  				& TexGS-VolVis 	   &9&0&1&0   \\
			\hdashline
			               	& IN2N  &2&2&3&3\\
 			beetle			& IG2G  &2&2&5&1      \\
							& DGE &0&2&2&6  \\
			  				& TexGS-VolVis 	   &6&4&0&0   \\
			\hdashline 
			               	& IN2N  &1&7&2&0\\
 			engine  		& IG2G  &1&1&2&6      \\
							& DGE &1&1&6&2  \\
			  				& TexGS-VolVis 	   &7&1&0&2   \\
			\hdashline 
			               	& IN2N      &2&1&5&2\\
 			ionization 		& IG2G  &0&1&4&5       \\
							& DGE &3&4&0&3  \\
			  				& TexGS-VolVis 	   &5&4&1&0  \\

		\end{tabular}
	}
	\label{tab:user-study-votes}
\end{table}


\vspace{-0.05in}
\bibliographystyle{abbrv-doi-hyperref}
\bibliography{template}

\end{document}